\newif\ifraw   % Schalter f"ur \tt-Stil (true schaltet auf \tt-Stil
\rawfalse
\documentclass[11pt,fleqn,leqno]{article}
\usepackage{amsmath}
\usepackage{amssymb}
\usepackage{amsthm}
\catcode`\@=11
\catcode`\@=12
\usepackage{enumerate}
\usepackage{xspace}
\newcommand{\zeile}{\hfill\\}
\newcommand{\komma}{\quad ,}
\newcommand{\punkt}{\quad .}
\swapnumbers
\theoremstyle{plain}
\newtheorem{theorem}{Theorem}[section]
\newtheorem*{theorem*}{Theorem}
\newtheorem{proposition}[theorem]{Proposition}
\newtheorem*{proposition*}{Proposition}
\newtheorem{lemma}[theorem]{Lemma}
\newtheorem*{lemma*}{Lemma}
\newtheorem{corollary}[theorem]{Corollary}
\newtheorem*{corollary*}{Corollary}
\theoremstyle{definition}
\newtheorem{definition}[theorem]{Definition}
\newtheorem*{definition*}{Definition}
\theoremstyle{remark}

\newtheorem*{remark*}{Remark}

\newtheorem*{note*}{Note}

\newtheorem*{cit*}{Citation}
\newcommand{\ein}{\hangindent1cm\hangafter1\zeile}

\newcommand{\RR}{\ensuremath{\mathbb{R}}}   % Symbol für reelle Zahlen
\renewcommand{\d}{\partial} % partieller Differentialoperator

\newcommand{\abs}[1]{\lvert#1\rvert}
\newcommand{\absbig}[1]{\left\lvert#1\right\rvert}

\newcommand{\const}{\ensuremath{\text{const}}}
\DeclareMathOperator{\tr}{tr}
\renewcommand{\div}{\ensuremath{\text{div}}\,}
\DeclareMathOperator{\grad}{grad}

\DeclareMathOperator{\vol}{Vol}

\DeclareMathOperator{\supp}{supp}

\newcommand{\leftsuper}[1]{{}^{\scriptscriptstyle #1}}

\newcommand{\vierGamma}{\leftsuper{4}\Gamma}
\newcommand{\viergamma}{\leftsuper{4}\gamma}
\newcommand{\viernabla}{\leftsuper{4}\nabla}
\newcommand{\vierR}{\leftsuper{4}\!R}
\newcommand{\viere}{\leftsuper{4}\!e}
\newcommand{\viersigma}{\leftsuper{4}\!\sigma}
\newcommand{\dd}[2]{\ensuremath{\partial_t^{#1}\partial_x^{#2}}}
\newcommand{\obar}[1]{\smash[t]{\overset{\rule[-0.5pt]{1ex}{0.5pt}}{#1}}}
\newcommand{\ubar}[1]{\smash[b]{\underset{\rule[5pt]{1ex}{0.5pt}}{#1}}}
\newcommand{\mass}{\ensuremath{\text{\sc m}}\xspace}
\ifraw
  \input wiggly.tex  % fuer wavy underlining
\fi
\title{Global Prescribed Mean Curvature foliations in cosmological
  spacetimes with matter\\ Part I}
\author{Oliver Henkel\thanks{Present address: Heinrich--Hertz--Institut
    f\"ur Nachrichtentechnik Berlin GmbH, Einsteinufer 37, 10587 Berlin,
    Germany}\\ 
  Max Planck Institute for Gravitational Physics\\ Am M\"uhlenberg 1\\
  14476 Golm, Germany}  
\date{October 18, 2001}
\begin{document}
\maketitle
\begin{abstract}
   This work investigates some global questions about cosmological
   spacetimes with two dimensional spherical, plane and hyperbolic symmetry
   containing matter.  
   The result is, that these spacetimes admit a global foliation by
   prescribed mean curvature surfaces, which extends at least 
   towards a crushing singularity. The time function of the foliation is
   geometrically defined and unique up to the choice of an initial Cauchy
   surface.\zeile
   This work generalizes a
   similar analysis on constant mean curvature foliations and avoids the
   topological obstructions arising from the existence problem.
\end{abstract}
\newpage
\tableofcontents
%
% Umschaltung auf \tt-Stil?
\ifraw
\renewcommand{\emph}{\underline}
\renewcommand{\textbf}{\underwiggle} 
\renewcommand{\baselinestretch}{2}
\tt
\fi
\section{Introduction}
The flavour of General Relativity viewed as an initial value problem
for the field equations is based on the geometrical nature of the
equations, which implies the diffeomorphism invariance and the absence of a
metric background structure. This has consequences for spacelike
foliations, in that the time function of a foliation is not canonical, but
arbitrary unless it is tied to some geometrical quantity. The latter would
turn the analysis of the global structure of spacetimes into an
investigation of the asymptotic behaviour of such a foliation.

In spacetimes with certain symmetries some global foliations, defined by
time coordinates tied to the symmetries are known, e.g. \cite{re},
\cite{an}.\zeile  
A geometrically defined foliation which does not depend on the symmetries
are the well known constant mean curvature (CMC) foliations, where the time
coordinate is given by the mean curvature of the leaves, which varies
continuously from leaf to leaf, see \cite{r1} for a survey of this topic and
\cite{r4}, \cite{r6} for foliations of spacetimes with symmetries.

Unfortunately the CMC constructions suffer from the existence problem,
which is unsolved in general. To overcome this difficulty, a foliation with
leaves of prescribed mean curvature (PMC) has been constructed 
at least locally in time in \cite{h2} for cosmological spacetimes. The
prescription is given implicitly, letting the mean curvature vary
continuously along the normal vector field of the foliation relative to a 
given Cauchy surface. The time function of the foliation is geometrically
defined and turns out to be intrinsic, in that coupled to Einstein's field 
equations adapted to the leaves, one obtaines Cauchy data for spacetimes
foliated by PMC leaves. 

The aim of the present paper is to globalize this result for certain
spacetimes. Motivated by the method in \cite{r3}, where satisfying results
about cosmic censorship have been obtained for spatially homogeneous models,
I consider here cosmological spacetimes with two dimensional spherical, plane
and hyperbolic symmetry. This choice is the first step of successively
lowering the degree of symmetry to obtain more and more general
results. I focus on the following questions:\zeile
How large is the maximal interval for the time function and does this
maximal foliation cover the whole spacetime?

One guideline taken from \cite{r3} is, that the foliation may be
extendible as long as the mean curvature of the leaves remains finite. We
will see, that this principle stays true with some modifications.\zeile
Another important aspect for the construction of global foliations is the
choice of an appropriate matter model. The key requirements for the matter
fields are their regularity in a regular geometric background as well as
some energy conditions. These requirements are
not trivially satisfied, since the first one rules out matter such as the
perfect fluid, which is known to 
develop singularities (shocks) in a regular geometric background. 
For dust it has been shown in \cite{r5}, \cite{ir}, that there is no way to
construct a global CMC foliation, which covers the whole dust-filled
cosmological spacetime. Thess counterexamples emphasize dramatically the
importance of choosing appropriate matter models. Furthermore demanding
energy conditions seems natural and obvious, but we will see, that the
non-negative pressures condition plays a special role in the improvements
in section \ref{s.improve}.

Section \ref{s.preliminaries} fixes notation and states some basic
definitions. We attack the main questions in section
\ref{s.surfacesym}, following closely the treatment in \cite{r4}.
In the final section the results will be discussed. 

% 
%
%
%
%

%%% Local Variables: 
%%% mode: latex
%%% TeX-master: "main"
%%% End: 

% 
%
%
%
%
%
\section{Basic definitions and formulas}\label{s.preliminaries}
\subsection{Spacetimes and foliations}\label{s.spacetimes}
A spacetime is a pair $(M,g)$, where $M$ denotes a four dimensional
smooth and orientable Lorentz manifold with metric $g$ and signature
$(-+++)$. The metric induces the Levi-Civita
connection $\viernabla$ on $M$. If $\{X_{\alpha}\}$ denotes a local basis of
vector fields on $M$, we define the connection coefficients
$\leftsuper{4}C_{\alpha\beta}^{\gamma}$ relative to this basis by 
$\viernabla_{X_{\alpha}}X_{\beta} = 
   \leftsuper{4}C_{\alpha\beta}^{\gamma} X_{\gamma}$
and we get the 
Christoffel symbols $\vierGamma$ and Ricci rotation
coefficients $\viergamma$ by specializing to a
coordinate basis or an orthonormal frame, respectively.
The sign convention for the curvature is fixed by the definition 
$\vierR(X,Y)Z := \viernabla_X\viernabla_Y Z - \viernabla_Y\viernabla_X Z 
               - \viernabla_{[X,Y]} Z$, 
where $X$, $Y$ $Z$ are vector fields. The curvature tensor is then defined
as $\vierR(W,Z,X,Y) := g(W, \vierR(X,Y)Z)$ with Ricci tensor 
$\vierR_{\alpha\beta}=\vierR^{\mu}_{\alpha\mu\beta}$
and scalar curvature $\vierR = \vierR^{\mu}_{\mu}$, written in abstract
index notation of the Ricci calculus.\zeile 
Then the Einstein tensor reads
$G_{\alpha\beta} = \vierR_{\alpha\beta} 
                 - \tfrac{1}{2}\vierR g_{\alpha\beta}$ 
and the field equations are 
$G_{\alpha\beta} = 8\pi T_{\alpha\beta}$ or equivalently 
$\vierR_{\alpha\beta} = 8\pi 
     \left( T_{\alpha\beta} - \tfrac{1}{2}(\tr T)\,g_{\alpha\beta} \right)$,
where $T_{\alpha\beta}$ denotes the energy momentum tensor of the matter
fields. The matter quantities are the energy density 
$\rho := T_{\mu\nu}n^{\mu}n^{\nu}$, 
the momentum density 
$j_{\beta} := -T_{\mu\nu}n^{\mu}h^{\nu}_{\beta}$
and the stress tensor 
$S_{\alpha\beta} := T_{\mu\nu}h^{\mu}_{\alpha}h^{\nu}_{\beta}$
with respect to an observer, represented by a unit timelike vector $n$, 
where $h_{\alpha\beta} := g_{\alpha\beta} + n_{\alpha}n_{\beta}$ denotes the
orthogonal projector on $\{n\}^{\perp}$ in covariant notation.\zeile
Einstein's field equations in vacuum ($T_{\alpha\beta}=0$) have a well
posed Cauchy 
problem in harmonic coordinates, thus one obtains spacetimes as solutions
of Einstein's field equations with matter, 
whenever the equations describing the matter fields and the energy momentum
tensor couple to the field equations in harmonic coordinates, such that the
Cauchy problem remains well posed. We will see examples in subsection
\ref{s.matter} (compare \cite{w1},\cite{fr} for an introduction/analysis of
the Cauchy problem for Einstein's equations). 

In this work we confine ourselves to \emph{cosmological} solutions $(M,g)$
of Einstein's field equations. Due to \cite{ba} these are globally
hyperbolic and spatially compact spacetimes, where the Ricci 
tensor contracted twice with any timelike vector is non-negative (timelike
convergence condition). This last condition can be reexpressed in terms of
the matter variables as $\rho+\tr S \ge 0$ for any observer, which is the
strong energy condition.

Now let us pay attention to an additional structure. 
A foliation $\{S_t\}$, $t \in I \subset \RR$ ($I$ interval containing zero)
of (a part of) $(M,g)$ by spacelike hypersurfaces induces on 
each leaf the unit normal vector field $n$, the metric 
$h_{\alpha\beta} = g_{\alpha\beta} + n_{\alpha}n_{\beta}$, which also 
serves as orthogonal projection and the second fundamental form
$k_{\alpha\beta}:=-h_{\alpha}^{\mu}h_{\beta}^{\nu}\,\viernabla_{\mu}n_{\nu}$
(the definition of $k_{\alpha\beta}$ fixes the sign conventions used in this
work). The second fundamental form is a symmetric tensor, intrinsic to the
leaves of the foliation, and can also be written as the
Lie derivative of the 
3-metric $h$ with respect to the normal vector field, 
$k_{\alpha\beta}=-\tfrac{1}{2}{\cal L}_n h_{\alpha\beta}$.
The 3-metric determines further geometrical objects on the leaves,
such as the Levi-Civita connection $\nabla$, the Christoffel symbols
$\Gamma$, the Ricci rotation coefficients $\gamma$ and the curvature tensor
$R(\cdot)$. Tensors intrinsic to the leaves of the foliation will carry
Latin indices in the abstract index notation.\zeile 
The parameter $t$ of the foliation has timelike gradient and thus can be
regarded as (coordinate-) time. Given
local coordinates $(x^i)$ on $S_0$, we can Lie-transport them to
neighbouring leaves along an arbitrary family of
transversal curves, parametrized by $x$. We will express equations
containing coordinate components with respect to the adapted
coordinates $(t,x)$.\zeile  
The lapse function $N$ and the shift vector $\nu \perp n$ on
the leaves are defined by
$\d_t = Nn + \nu$, thus 
$N = -g(\d_t,n)$ and $\nu = \d_t - Nn$.
Then we have $1 = dt(\d_t) = N\,dt(n)$. Further, $dt$ is
(co-)orthogonal to the leaves and if we denote the conormal of
the leaves by $\sigma$ we see that 
$dt = -N^{-1}\sigma$ or $\sigma = -N dt$,
thus $N^{-1}$ measures the length of $dt$.\zeile
The most common example involving the lapse function is the event
horizon in Schwarz\-schild spacetime, 
where the coordinate time explodes along the worldline of infalling
observers $n$, thus $N^{-1}$ explodes or $N \longrightarrow 0$. In
this work we will be faced with a complementary scenario, where we
have to ensure, that the lapse function does not explode,
corresponding to the phenomen of recollapse, where coordinate time
freezes, as $dt \longrightarrow 0$. If this occurs, one could try to
reparametrize the foliation by setting $\tilde{t}=f(t)$ for some
monotone function $f$ and one gets $d\tilde{t}=f'(t)\,dt$,  
\begin{equation*}
   \tilde{N}=\tfrac{1}{f'(t)}N=\tfrac{dt}{d\tilde{t}}N 
   \komma 
\end{equation*} 
and the same relation holds for shift.

The $3+1$ split of the spacetime geometry by means of lapse and shift ends
up in the $3+1$ form of the field equations. The constraint equations are 
\begin{subequations}\label{e.constraints}
\begin{gather}
   R + H^2 - \abs{k}^2 = 16 \pi \rho 
   \qquad\,\text{(Hamiltonian constraint)}\\
   \nabla^j k_{ij} - \nabla_i H = 8\pi j_i
   \qquad\quad\!\text{(momentum constraint)}
   \komma
\end{gather}
\end{subequations}
with $\abs{k}^2 = k_{\alpha\beta}k^{\alpha\beta}$ and $H=\tr k$
denotes the mean curvature of the leaves. The ADM equations read
\begin{subequations}\label{e.adm}
\begin{gather}
   \d_t h_{ij} = -2Nk_{ij} + \nabla_i\nu_j + \nabla_j\nu_i \\
   \d_t k_{ij} = -\nabla_i\nabla_j N
                   + N \Big(
                         R_{ij} + Hk_{ij} - 2k_i^r k_{rj} -8\pi
                         (\, S_{ij}+\tfrac{1}{2}(\rho-\tr S)h_{ij} \,)
                      \Big) \\
   \qquad\quad\,\, + \nu^r\nabla_r k_{ij} 
                   + k_{rj}\nabla_i\nu^r + k_{ir}\nabla_j\nu^r 
                   \notag
   \punkt
\end{gather}
\end{subequations}
Taking the
trace of the second equation and eliminating the scalar curvature $R$ by
the Hamiltonian constraint we obtain the lapse equation
\begin{equation}\label{e.lapse}
   \Delta N + N \Big(
                   \abs{k}^2 + 4\pi (\, \rho + \tr S \,)
                \Big)
     = (\d_t -\nu) H
   \komma
\end{equation}
which serves as a constraint of the foliation. Note, that in cosmological
spacetimes, the term in brackets is 
always non-negative. If it turns out to be non-vanishing, then the
left-hand side of \eqref{e.lapse} can be shown to be an isomorphic mapping
of $N$, considered as an element of some Sobolev space $H^s$ into
$H^{s-2}$. This observation motivates the following 
\begin{definition}\label{def.pmc}\ein
   A \emph{Prescribed Mean Curvature (PMC) foliation} is defined
   to be a foliation $\{S_t\}$ satisfying \eqref{e.lapse}, with 
   \begin{equation}\label{e.pmc}
      (\d_t -\nu) H = Nn(H) := 1
      \komma 
   \end{equation} 
   thus the mean curvature of the leaves is forced to vary
   uniformly along the normals of the leaves.
\end{definition}

In \cite{h2} I proved the following local in time result: 
{\samepage
\begin{theorem}\label{thm.pmclocal}\ein
   Let $(M,g)$ be a smooth, globally hyperbolic spacetime, obeying the
   strong energy condition, with compact Cauchy surface $\Sigma$ and
   \begin{equation}\label{e.lambda}
      \lambda = \abs{k}^2 + 4 \pi (\rho + \tr S)
   \end{equation}
   does not vanish identically on $\Sigma$.\zeile
   Then there exists a $T>0$ and a unique smooth PMC foliation
   $\{S_t\}$,  $t \in [-T,T]$ in $(M,g)$, with $\Sigma=S_0$.
\end{theorem}
}

Note, that the setting here is quite general, no
symmetry assumptions have to be made and essentially the strong energy
condition turns out to be sufficient for the local in time existence of a
unique PMC foliation up to the choice of an initial Cauchy surface. 

The aim of the present paper is to globalize the result. Here are two problems
involved: How large is the interval of values taken by the time coordinate
and does the global foliation then cover the whole spacetime? To answer
these questions in general there seem to be no techniques available up to
now. One 
strategy to obtain global results is, to study first spacetimes with some
spatial symmetry, taking advantage of the simplifications of the
equations. Then the hope is, that the techniques developed in these cases
give insight into the nature of more general classes of spacetimes, by
successively lowering the degree of symmetry. Here we will focus on spacetimes
with two dimensional spacelike orbits of symmetry and three (local) Killing
vector fields. In the second part of this paper we will consider spacetimes
with two commuting (local) Killing vector fields. These cases are indeed
the first steps of this program, since the initial analysis for locally
spatially homogeneous spacetimes has already been done and lead to very
strong results (see \cite{r3} for
the exact analysis or \cite{diss} for an overview about the results in the
present context). 
\subsection{Matter models}\label{s.matter}
Before getting deeper into the analysis just motivated I
introduce some matter models and their coupling to 
the Einstein equations, with special emphasis on the Cauchy problem. Since
energy conditions will play an important role 
in the estimates we perform later on, I assemble the relevant ones for the
present work first. 
\begin{itemize}
\item The dominant energy condition. Its statement is,
   that for all orthonormal frames $\{\viere_{\alpha}\}$ with $\viere_0$
   timelike 
   \begin{equation*}
      T(\viere_0,\viere_0) \ge \abs{T(\viere_{\alpha},\viere_{\beta})}
      \punkt
   \end{equation*}
   Another
   formulation of this condition is, that for any observer,
   the local energy density $\rho$ is
   non-negative and the momentum density 
   $j$ is non-spacelike, thus 'matter cannot travel faster than light', a
   statement, that can be proved rigorously (see for example section 4.3 of
   \cite{he}), leading to the result, that if the 
   energy momentum tensor obeys the dominant energy condition and vanishes
   on a set $S$, then it vanishes on the whole Cauchy development $D(S)$ of
   the set, thus $D(S)$ is a vacuum spacetime.
\item The strong energy condition states that for all
   timelike vectors $v$ the inequality  
   \begin{equation*}
      T(v,v) \ge \tfrac{1}{2}(\tr T)g(v,v)
   \end{equation*}
   holds, or equivalently for any observer $\rho+\tr S \ge 0$, thus the
   stresses do not become too negative. Another
   formulation is $\vierR(v,v) \ge 0$, also known as the timelike
   convergence condition, which contributes to the expansion of timelike
   geodesic congruences a negative term, thus shifting the balance towards
   contraction to a final singularity of the congruence. Hawking's famous
   singularity theorems guarantees then geodesic incompleteness in the past
   provided further the existence of a Cauchy surface with
   uniform negative mean curvature.\zeile
   We already used the strong energy condition as an integral part in the
   definition of cosmological spacetimes and its meaning for foliations.
\item The non-negative pressures condition 
   demands the stress tensor $S$ to be positive definite. This 
   condition ensures in some sense, that the pressures contribute more to
   attraction than to repulsion, leading to a finite lifetime of the
   spacetime under certain circumstances.
   This somewhat unexpected behaviour is a
   true relativistic effect, which does not have a Newtonian counterpart.
\end{itemize}
\subsubsection{The Maxwell equations}
The Maxwell field is described
by a two form $F$, subject to the Maxwell equations
\begin{subequations}\label{e.maxwell}
\begin{gather}
   dF = 0 \qquad\quad\iff
   \viernabla_{[\kappa}F_{\mu\nu]} = 0 \\
   d\star F= 4\pi \star J \iff
   \viernabla_{\mu}F^{\alpha\mu} = 4\pi J^{\alpha}
   \komma
\end{gather}
\end{subequations}
where $J$ denotes the electromagnetic charge current density. Alternatively
the covariant derivatives can be replaced by ordinary derivatives.\zeile
It is well known, that the Maxwell equations admit (locally) a
reformulation in terms of the vector potential $A$, a one form with 
$dA = F$. Then the remaining inhomogeneous equation reads
$\viernabla^{\mu}\viernabla_{\alpha}A_{\mu} 
 - \viernabla^{\mu}\viernabla_{\mu}A_{\alpha} = 4\pi J_{\alpha}$. Fixing the
gauge invariance of the equation by the Lorentz gauge, we obtain the system
\begin{subequations}\label{e.maxwell-lorentz}
\begin{gather}
   \viernabla^{\mu}A_{\mu} = 0 \qquad \text{(Lorentz-gauge)} \\
   \square A_{\alpha} = 4\pi J_{\alpha} - R_{\alpha\mu}A^{\mu}
   \komma
\end{gather}
\end{subequations}
with $\square = - \viernabla^{\mu}\viernabla_{\mu}$ and the curvature term
arises as a consequence of some commutations of derivative operators.
In this formulation the second equation is a wave equation, hence can be
brought into first order symmetric hyperbolic form, if the source term is
appropriate. Thus we get local 
existence and uniqueness for this equation in a given spacetime. One
can further show, 
that then the Lorentz gauge propagates and we get indeed a unique local
solution of the Maxwell equations. Spatially global solution can be obtained
by the usual patching argument, arising from localizing the equation with
respect to an appropriate partition of unity of the initial data on some
Cauchy surface. 

Given an electromagnetic field $F$ we can form the associated energy
momentum tensor $E$, defined as
\begin{equation}\label{e.maxwellenergy}
   E_{\alpha\beta} = \frac{1}{4\pi} 
     \left( F_{\alpha\mu}{F_{\beta}}^{\mu} 
          - \tfrac{1}{4} \abs{F}^2\, g_{\alpha\beta} 
     \right)
   \komma
\end{equation}
where $\abs{F}^2 = F^{\mu\nu}F_{\mu\nu}$.\zeile
$E$ is tracefree, satisfies the relation 
$\viernabla_{\mu} E^{\alpha\mu} = -{F^{\alpha}}_{\mu} J^{\mu}$ 
and the dominant and strong energy condition hold. 

Since the Maxwell equations turn out to be symmetric hyperbolic at least
locally, we can couple them to the Einstein equation to get a symmetric
hyperbolic system of equations in harmonic coordinates. Thus we end up with
a well-posed Cauchy problem for the Einstein-Maxwell system, as long as the
electromagnetic charge current density is in appropriate form.
\subsubsection{The Vlasov equation}
The Vlasov equation is a model for a collisionless gas. It describes the 
motion of a huge number of structureless particles in spacetime. We need
only the case, where the particles have unit mass, where the equation is
composed of a non negative function $f$, defined on the mass shell of
particles of mass one 
$P := \{v \in TM \;|\; g(v,v)=-1 \text{, future pointing}\}$,
representing the particle distribution, and a
geodesic spray $X$ on $P$. The equation reads
\begin{subequations}\label{e.vlasov}
\begin{equation}
   X(f) = 0
   \komma
\end{equation}
with
\begin{equation}
   X = p^{\mu}\d_{\mu} - \vierGamma_{\mu\nu}^{k}\, 
                         p^{\mu}p^{\nu} \frac{\d}{\d p^{k}} 
     = v^{\alpha} \viere_{\alpha} - \viergamma_{\mu\nu}^{k}\, 
                               v^{\mu}v^{\nu} \frac{\d}{\d v^{k}}
   \komma
\end{equation}
\end{subequations}
where $p^{\mu}$ denotes the components of the momentum of the particles
with respect to the given coordinates and $v^{\alpha}$ are the 
components of the momentum with respect to an orthonormal frame. They are
related by $p^{\mu} = (\viere_\alpha)^{\mu}\, v^{\alpha}$ and on the mass
shell $p^0$ is determined by the components $p^i$ and 
$v^0 = \sqrt{1+\delta_{ij}v^iv^j}$.\zeile
Inserting the definition
$\viere_{\alpha}={(\viere_{\alpha})}^{\mu}\d_{\mu}$ and doing a 3+1
decomposition (with the exception, that we do not write down the explicit
expression for $\viergamma_{0j}^k$) we can reformulate the
Vlasov equation as
\begin{equation}\tag{\ref{e.vlasov}'}\begin{split}
   0 = \d_t f & + (N\tfrac{v^j}{v^0}({e_j})^i-\nu^i)\d_i f \\
       & -\Big(
            e_i(N)v^0 
          + N(-k_{rs}({e_i})^r({e_j})^s+\delta_{ik}\viergamma_{0j}^k) v^j
          + N\delta_{ik}\gamma_{rs}^k\tfrac{v^r v^s}{v^0}
          \Big)
          \frac{\d}{\d v^i} f
   \punkt
\end{split}\end{equation}

The Cauchy problem for the Vlasov equation in a given spacetime is
easy, since it is a linear, scalar equation with characteristic vector
field $(Y,Q)$ satisfying 
\begin{gather*}
   \dot{Y}^{\alpha} = Q^{\alpha} \\
   \dot{Q}^{\alpha} = -\vierGamma_{\kappa\lambda}^{\alpha}
                       Q^{\kappa}Q^{\lambda}
   \komma
\end{gather*}
and since $X(f)=0$, $f$ is constant along the characteristics. 

From the particle distribution $f$ one can construct other physically
meaningful quantities such as the energy momentum tensor by integration
over the tangent spaces. We denote the part of 
the mass shell in the fibre over $x \in M$ by $P_x := P \cap T_xM$.
Then we define the energy-momentum tensor $T$ by 
\begin{equation}\label{e.vlasovenergy}
   T_{\alpha\beta}(x) := - \int_{P_x} f p_{\alpha}p_{\beta} 
                                      \sqrt{\abs{g}}/p_0\,dp^1dp^2dp^3
   \punkt
\end{equation}
$T$ is divergence free, satisfies the dominant and strong energy condition
and the non-negative pressures condition.\zeile
To obtain the matter quantities $\rho$, $j$ and $S$ one has to 
calculate the components of the energy momentum tensor with respect to an
orthonormal frame. These components have the following representation:
\begin{equation}\tag{\ref{e.vlasovenergy}'}
   T(\viere_{\alpha},\viere_{\beta})(x) 
     = \int_{P_x} f v_{\alpha}v_{\beta}/v^0\,dv^1dv^2dv^3
\end{equation}

Now let us consider the Einstein-Vlasov system. The energy momentum
tensor automatically satisfies $\div T=0$ as mentioned above, thus
there arise no additional equations from the Bianchi identities. 
But unfortunately the coupled system of equations is not symmetric
hyperbolic in any sense, due to the fact, that it is a system of
integro-differential equations. Nevertheless, the local existence proof
applies similar techniques as in 
the case of quasilinear symmetric hyperbolic systems where the
peculiarity in the construction for the Einstein-Vlasov system
consists in bounding the support of $f$ in the tangent space: There
is no a priori bound on the velocities of the particles, and no
localization argument available as for the spacetime coordinates. In
order to estimate the matter quantities appearing in the coupled system,
one has to control the maximal velocity uniformly during the
construction. This has been done for example in \cite{r2}, establishing a
well-posed Cauchy problem for the Einstein-Vlasov system.\zeile 
With this result at hand, it is easy to extend this result to the
Einstein-Vlasov-Maxwell  
system, since the Einstein-Maxwell equations are symmetric hyperbolic
and when coupling the Vlasov equation to them nothing new appears.
\subsection{Further conventions}
For the convenience of the reader I cite the well known Gronwall estimate,
which is central to the analysis of partial differential equations and will
be used here in this work. I adopt the convention of denoting
any generic constant by $C$. 

\begin{proposition}[Gronwall's inequality]\label{prop.gronwall}\ein
   Let $I \subset \RR$ be an interval, $t_0 \in I$ and
   $\alpha, \beta, u \in C(I,\RR_+)$, with
   \begin{equation*}
      u(t) \le \alpha(t) 
           + \absbig{\int_{t_0}^t \beta(s)u(s)\,ds }
   \end{equation*}
   for all $t \in I$.\zeile
   Then 
   \begin{equation*}
      u(t) \le \alpha(t)
           + \absbig{ \int_{t_0}^t \alpha(s)\beta(s)\:
                      e^{\abs{\int_s^t \beta(r)\,dr}}\: ds}
   \end{equation*}
   holds for all $t \in I$.
\end{proposition}
The proof of Gronwall's inequality in this particular form can be found 
in \cite{a}. 
%
%
%
%
%
%
%
%
%
%
%

%%% Local Variables: 
%%% mode: latex
%%% TeX-master: "main"
%%% End: 

%
%
%
%
%
%
\section[Spacetimes with spherical, plane and hyperbolic
symmetry]{Spacetimes with two dimensional spherical, plane and hyperbolic
  symmetry}
\label{s.surfacesym}
\subsection{The geometry of surface symmetric spacetimes} \label{s.geometry}
Let $(M,g)$ be a smooth, globally hyperbolic spacetime which is
topologically of 
the form $\RR \times S^1 \times F$, with $F$ a compact, orientable
surface. The submanifolds $\{\tau\} \times S^1 \times F$ are assumed to
be Cauchy surfaces of $M$. The universal covering $\hat{F}$ of $F$
induces a spacetime 
$(\hat{M},\hat{g})$ by $\hat{M}=\RR \times S^1 \times \hat{F}$
and $\hat{g}=p^*g$, $p: \hat{M }\longrightarrow M$ the canonical
projection. Moreover, there is a group $G$ of isometries acting on
$(\hat{M},\hat{g})$.\zeile 
Then $(M,g)$
\begin{itemize}
\item is called \emph{spherically symmetric}, if $F=S^2$ and $G=SO(3)$ acts
   isometrically and without fixed points on $S^1 \times S^2$
\item is \emph{plane symmetric}, if $F=T^2$ and $G=E_2$ (Euclidian
   group) acts isometrically on $\hat{F}=\RR^2$
\item has \emph{hyperbolic symmetry}, if $F$ has genus greater than one
   and the connected component of the symmetry group $G$ of the hyperbolic
   plane, $H^2$, acts isometrically on $\hat{F}=H^2$ (thus
   $F=H^2/\Gamma$, with $\Gamma$ a discrete group of isometries of $H^2$)
\end{itemize}
and the matter quantities remain invariant under the isometries.\zeile
To collect these cases, each such spacetime is called \emph{surface
  symmetric}, the diffeomorphic images of $F$ in the product
decomposition of $M$ \emph{surfaces of symmetry} and each surface in $M$
diffeomorphic to $S^1 \times F$ will be called \emph{symmetric}.\zeile
In expressions involving indices, lower case Greek indices range from
0 to 3, lower case Latin indices (preferably taken from the middle of the
alphabet) range from 1 to 3 and upper case Latin indices (from the
beginning of the alphabet) take the values 2 or 3.

The isometric action forces the curvature of the surfaces of symmetry up to
rescaling to be $\epsilon=1,0,-1$ in the spherical, plane and hyperbolic
case, respectively. Therefore they can be coordinatized by the well known
angles $(\vartheta,\varphi)$ which cast the metric 
$\tilde{g}$ of the surfaces of symmetry (considered for a moment as abstract
manifolds) into the form
\begin{equation}\label{e.gtilde}
   \tilde{g} = d\vartheta^2 + \epsilon_{\vartheta}^2\, d\varphi^2
   \komma\quad
   \epsilon_{\vartheta} := \left\{ \begin{matrix} 
                        \sin\vartheta  &        & \epsilon=1 \\ 
                        1              & \komma & \epsilon=0 \\ 
                        \sinh\vartheta &        & \epsilon=-1
                     \end{matrix} \right.
   \punkt
\end{equation}
Define the area radius function $r$ on a surface of symmetry $F$ (embedded in
$M$) to be 
\begin{equation}\label{e.radius}
   r=\sqrt{\frac{1}{4\pi}\vol(F)}
   \komma
\end{equation}
then $r$ is independent of $(\vartheta,\varphi)$ and the metric of $F$
reads
\begin{equation}\label{e.gobar}
   \obar{g} = r^2 \tilde{g}
   \punkt
\end{equation}
With respect to any symmetric Cauchy surface $S$ we have
the timelike unit normal vector $n$ of $S$ in $M$. Regarding $n$ as a normal
vector to $F$ in $M$, we can define a second unit normal $m$ of $F$ by the
conditions, that $m$ is tangent to $S$ and the system
$(n,m,\obar{e}_2,\obar{e}_3)$ is positively oriented, where we set
$\obar{e}_A:=r^{-1}\tilde{e}_A$, $\tilde{e}_2:=\d_2$, 
$\tilde{e}_3:=\epsilon_{\vartheta}^{-1}\d_3$. 
We denote the
associated second fundamental forms of $(F,\obar{g})$ in $(M,g)$ by
$\kappa$ and $\lambda$, with
\begin{subequations}\label{e.lambdakappa}
\begin{gather}
   \lambda_{AB} = -\tfrac{1}{2}m(\obar{g}_{AB})
                = -\tfrac{1}{2}m(r^2)\,\tilde{g}_{AB}
                = \tfrac{1}{2} (\tr\lambda)\, \obar{g}_{AB} 
   \komma\quad \tr\lambda = -\tfrac{2}{r}m(r) \\
   \kappa_{AB}  = -\tfrac{1}{2}n(\obar{g}_{AB})
                = -\tfrac{1}{2}n(r^2)\,\tilde{g}_{AB}
                = \tfrac{1}{2} (\tr\kappa)\, \obar{g}_{AB} 
   \komma\:\:\,\quad \tr\kappa = -\tfrac{2}{r}n(r) 
\end{gather}
\end{subequations}

Consider now a Gaussian coordinate neighbourhood
$(x',\vartheta,\varphi)$ of a surface of symmetry $F$, covering
(a part of) a symmetric Cauchy surface $S$. The metric $h$ of $S$ then
takes the form $h = d{x'}^2 + \sqrt{\abs{h}} \tilde{g}$. The
projection of geodesics starting in $(\hat{M},\hat{g})$ orthogonal at
$F$ remain orthogonal to all surfaces of symmetry. Following
them until their projection meets $F$ again, the symmetry allows only
two possibilities: The point of return 
is the same as the starting point or an antipodal point, in which case
we force the geodesic to turn a second time around the circle. Let $L$
denote the length of the geodesic. Setting 
$a=2\pi \left( \int_0^L \abs{h(z)}^{-1/4}\,dz \right)^{-1}$, we define a
new coordinate $x(x')$ by $a \int_0^{x'} \abs{h(z)}^{-1/4}\,dz$. In the
coordinates $(x^i)=(x,\vartheta,\varphi)$ the metric has the representation
\begin{equation}\label{e.h}
   h = A^2 
       \left(
          dx^2 + a^2 \tilde{g}
       \right)
     = A^2 \,dx^2 + \obar{g}
   \komma
\end{equation}
with $A(x)=a^{-1}\abs{h(x)}^{1/4}$ defined on $S^1$. Comparing this with
equation \eqref{e.gobar} shows $r=Aa$.\zeile
The corresponding Laplacian $\Delta$ acting on a function $\psi$ on $S$ can
now be calculated to 
\begin{align*}
   \Delta\psi & = - h^{ij}\nabla_i\nabla_j\psi
                = - h^{11}\nabla_1\nabla_1\psi - h^{AB}\nabla_A\nabla_B\psi
                \\
              & = - h^{11}\nabla_1\nabla_1\psi 
                  + h^{AB}\Gamma_{AB}^1\psi' 
                  + \obar{\Delta}\psi \\
              & = -A^{-2}(\psi''+A^{-1}A'\psi') + \obar{\Delta}\psi
   \komma
\end{align*}
where $\obar{\Delta}$ denotes the Laplacian of $\obar{g}$, and the prime
differentiation with respect to $x$, a convention, which we adopt for the
rest of this work.\zeile
Furthermore, the symmetry and the given coordinate representations permits 
the second fundamental form $k$ to have the form
\begin{equation}
   \label{e.k}
   k = A^2 K\, dx^2 + \tfrac{1}{2}(\tr\kappa) \obar{g} 
   \komma
\end{equation}
where the coefficents are functions on $S^1$. Taking the trace yields the
mean curvature $H=\tr k$ and we get the relation 
\begin{equation}\label{e.trkappa}
   H-K = \tr\kappa
   \punkt
\end{equation}

So far we know the intrinsic and extrinsic geometry for a symmetric
Cauchy surface $\Sigma$ in $M$. Let us turn now to the 3+1-geometry.
Theorem \ref{thm.pmclocal} states the conditions to guarantee local in time 
existence for a PMC foliation $\{S_t\}$ of a neighbourhood of
$\Sigma$. We need, that $\lambda$ defined by equation \eqref{e.lambda}
is non-negative and does not vanish identically on $\Sigma$. Again $n$ is
the unit normal on $\Sigma$ and $T$ the energy-momentum tensor of 
the spacetime. Therefore we get a sufficient condition by the assumption
that $M$ satisfies the strong energy condition (hence the matter term in
\eqref{e.lambda} is non-negative) and that
there exists at least one point in $\Sigma$, with $\lambda>0$. The
strictness of the inequality is not a restriction, since one always can
perform a slight deformation of $\Sigma$, such that the second
fundamental form is not identically zero, which does the job. The same
reasoning works, of course, if $M$ satisfies the dominant energy condition
and the non-negative-pressures condition (then both parts of the matter
term in \eqref{e.lambda} are non-negative). 
Thus, in surface symmetric spacetimes, some energy conditions are
sufficient for the existence of a local in time PMC foliation.

Having constructed a local in time PMC foliation, one can ask,
if the leaves of the foliation are symmetric, when
$\Sigma$ has been chosen to be symmetric (Note, that due to
symmetry $\lambda=\lambda(x)$ is a function on $S^1$ only).
As described in \cite{h2} the PMC foliation is given as a limit
$(w,N)$ of functions $(w^j,N^j)$ on $\Sigma$ 
in some Sobolev space. Here $w^j$ describes a family of spacelike
hypersurfaces in $M$ and $N^j$ converges towards the lapse function of
$w$. $(w^j,N^j)$ are defined as solutions of the sequence of 
symmetric--hyperbolic elliptic systems 
\begin{gather*}
   \d_t w^j + A^i(w^{j-1},N^{j-1})\,\d_i w^j 
       + B(w^{j-1},N^{j-1},DN^{j-1}) = 0 \\
   \Delta(w^{j-1})N^j + \lambda(w^{j-1},Dw^{j-1}) N^j = 1
   \komma
\end{gather*}
with $w^0$ representing $\Sigma$ and $N^0=1$, hence respecting the
symmetries. The underlying metric structure of the system is the sequence
of first fundamental forms $h_{j-1}$ of the surfaces $w^{j-1}$. If all the
$(w^j,N^j)$ respect the symmetries, then the PMC foliation also, hence it
suffices to show, that $(w^j,N^j)$ respects the symmetries, if
$w^{j-1},N^{j-1}$ does so.

This is clear for $w^j$, because the symmetric hyperbolic equation can be
localized, and the pullback to $\hat{M}$ is invariant under the
action of the isometry group.\zeile
Suppressing the index $j-1$ from some quantities determined by the metric
$h_{j-1}$, the Laplacian reads 
\begin{equation*}
   \Delta(w^{j-1}) N^j 
      = -A^{-2} \left( (N^j)'' + A^{-1}A'(N^j)' \right) +
      \obar{\Delta}(w^{j-1}) N^j 
   \punkt
\end{equation*}
The elliptic equation then  
becomes
\begin{equation*}
   -A^{-2} \left( (N^j)'' + A^{-1}A'(N^j)' \right) 
   + \lambda(w^{j-1},Dw^{j-1}) N^j
   + \obar{\Delta}(w^{j-1}) N^j
   = 1
   \punkt
\end{equation*}
If $\obar{\Omega}=\sqrt{\abs{\obar{g}}}=r^2\epsilon_{\vartheta}$ denotes
the volume form of the surfaces of symmetry $F$ in $M$, one finds 
$\vol(F)=\int_{[0,\pi]\times[0,2\pi]} \obar{\Omega}=4\pi r^2$, 
as desired.
Setting $L := -A^{-2} ( \frac{d^2}{dx^2} + A^{-1}A'\frac{d}{dx} )$ and
$\tilde{N}^j(x) := \frac{1}{\vol(F)} \int_{[0,\pi]\times[0,2\pi]} N^j
                   \obar{\Omega} 
                 = \frac{1}{4\pi} \int_{[0,\pi]\times[0,2\pi]} N^j
                   \epsilon_{\vartheta}$,  
then integration of the elliptic equation,
$\int_{[0,\pi]\times[0,2\pi]} 
   \epsilon_{\vartheta} (L+\lambda+\obar{\Delta})N^j 
 = \int_{[0,\pi]\times[0,2\pi]} \epsilon_{\vartheta} = 4\pi$, 
yields 
\begin{equation*}
   (L+\lambda+\obar{\Delta})\tilde{N}^j = 1
   \komma
\end{equation*}
since on the one hand 
$\int_{[0,\pi]\times[0,2\pi]} 
   (L+\lambda)N^j\epsilon_{\vartheta}=4\pi(L+\lambda)\tilde{N}^j$ 
and on the other hand 
$\int_{[0,\pi]\times[0,2\pi]} (\obar{\Delta}N^j)\epsilon_{\vartheta}
 = r^{-2}\int_{[0,\pi]\times[0,2\pi]}
   (\obar{\Delta}N^j)\obar{\Omega}=r^{-2}\int_F
   \obar{\Delta}N^j$ 
vanishes as well as $\obar{\Delta} \tilde{N}^j$.\zeile
Thus $\tilde{N}^j$ is a solution of the elliptic equation. Uniqueness then
gives us  $N^j = \tilde{N}^j$, thus $N^j=N^j(x)$, which 
expresses the symmetry of $N^j$, as desired.

Thus we have shown, given a symmetric Cauchy surface $\Sigma$ in $M$
admitting a local in time PMC foliation, that all leaves of the foliation
are symmetric, too and coordinatizing them in the way described above
yields: 
\begin{proposition}\label{prop.4geometry}\ein
   Let $(M,g)$ be a surface symmetric spacetime obeying the 
   strong energy condition.\zeile
   Then there are coordinates $(x^{\mu})=(t,x,\vartheta,\varphi)$ adapted
   to a local in time PMC foliation $\{S_t\}$ of a neighbourhood
   $U=]t_1,t_2[\, \times \Sigma$ of $\Sigma=S_0$ in $M$, which cast the 
   metric into the form
   \begin{equation}\label{e.g}
      g = -N^2 dt^2 + A^2 \left( (dx + \nu dt)^2 + a^2 \tilde{g}
      \right)
      \punkt
   \end{equation}
   All coefficients except $a$ are functions on $]t_1,t_2[ \,\times\, S^1$,
   whereas $a$ depends only on the time function $t$. $A$ and $a$ are
   everywhere positive, $N$ 
   denotes the lapse function of the foliation and $\nu$ the non-vanishing
   component of the shift vector, uniquely fixed by the condition 
   $\nu(t,0)=0$.
\end{proposition}

Supplementary to the notation already introduced, let an overdot denote
differentiation with respect to $t$, while keeping the prime as a marker
for differentiation with respect to the coordinate $x$.

Finally we find for the orthonormal frame $\{\viere_{\mu}\}$ and its dual 
$\{\viersigma^{\mu}\}$ canonically
induced by the $3+1$--split:
\begin{equation*}
   \begin{aligned}
      \viere_0 & = n = N^{-1}(\d_0 - \nu \d_1) \\
      \viere_1 & = m = A^{-1} \d_1 \\
      \viere_2 & = (Aa)^{-1} \d_2 \\
      \viere_3 & = (Aa\epsilon_{\vartheta})^{-1} \d_3
   \end{aligned}
   \qquad\qquad
   \begin{aligned}
      \viersigma^0 & = N dt \\
      \viersigma^1 & = A (\nu dt + dx^1) \\
      \viersigma^2 & = Aa\, dx^2 \\
      \viersigma^3 & = Aa\epsilon_{\vartheta}\, dx^3
   \end{aligned}
   \punkt
\end{equation*}
%
%
%
%
%
%
%
%
%
%
%\pagebreak
\subsection{The f\/ield equations}\label{s.fieldequations}
Given the PMC foliation, we can write down the field equations in the
3+1-representation. The symmetries suggest to represent the 
matter quantities completely with respect to an
orthonormal frame, thus we define
$j:=-T(n,m)=A^{-1}j_1$ and 
$S_{ij}=T(\viere_i,\viere_j)$.\zeile 
Then one calculates the constraint equations \eqref{e.constraints} to be
\begin{gather}
   \label{e.ch}
   (A^{1/2})^{\prime\prime} = \tfrac{1}{8}A^{5/2}
      \Big( H^2 - \tfrac{1}{2}(H-K)^2 - K^2 - 16\pi\rho \Big)
      + \tfrac{1}{4}A^{1/2}a^{-2}\epsilon \\
   \label{e.cm}
   K' = -3 A^{-1}A'K + A^{-1}A'H + H' + 8\pi Aj \\
   \intertext{The foliation is fixed by the lapse equation \eqref{e.lapse}
     and the PMC condition \eqref{e.pmc}} 
   \label{e.l}
   N^{\prime\prime} = -A^{-1}A'N' + A^2N 
      \Big( \tfrac{1}{2}(H-K)^2 + K^2
             + 4\pi(\rho+\tr S) 
      \Big) -A^2  \\
   \label{e.p}
   \dot{H} = 1 + \nu H' \\
   \intertext{and the evolution equations \eqref{e.adm} read} 
   \label{e.e1}
   \dot{A} = -NAK + A\nu' + A'\nu \\
   \label{e.e2}
   \dot{a} = -\tfrac{1}{2}Na(H-3K) - a\nu' \\
   \label{e.e3}
   \dot{K} = \nu K' - A^{-2}(N'' - A^{-1}A'N') \\\notag
      \qquad + N\left( 
          -4 A^{-5/2} (A^{1/2})'' + (A^{-2}A')^2 + HK
          -8\pi(A^{-2}S_{11}+\tfrac{1}{2}(\rho-\tr S)) 
      \right)
\end{gather}
Integrating the equation for $a$ over the circle yields
$\dot{a}=\mu a$, with $\mu=-\tfrac{1}{2} \int_{S^1}N(H-3K)$,
since $\int_{S^1}\nu'$ vanishes. Inserting this back gives an equation
for shift:
\begin{equation}\label{e.s}
   \nu' = -\tfrac{1}{2}N(H-3K) + \tfrac{1}{2}\int_{S^1}N(H-3K)
\end{equation}
Differentiation of the equation for $H$ with respect to $x$ yields an
equation for $H'$:
\begin{equation}\label{e.p'}
   (\d_t - \nu \d_x)H' = \nu' H'
\end{equation}
In summary we have equations for space and time derivatives of the
fundamental forms $h$ and $k$. Moreover there are equations for the
spacelike derivatives of lapse and shift, but unfortunately there is
no information about their time derivatives.

So far we started with a given surface symmetric spacetime,
admitting a local in time PMC foliation, to which we adapted the field
equations. This raises the opposite question: Given the equations
\eqref{e.ch}-\eqref{e.p'} and appropriate data, does there exist a
solution, and how unique is it? To answer this question we
first need to state more precisely the term 'appropriate data':
\begin{definition}\label{def.admdata}\ein
   A {\em symmetric initial data set} is a smooth collection $(\Sigma,h,k)$
   consisting of a 3-manifold $\Sigma$ diffeomorphic to $S^1 \times F$ with
   metric 
   $h$ and a symmetric tensor field $k$ on $\Sigma$, where $\Sigma$
   admits coordinates, such that $h$ and $k$ could be written in the form
   shown in section~\ref{s.geometry}.\zeile
   If there are matter fields present, then it is assumed, that
   there is also smooth symmetric matter data and equations, leading to a
   well posed Cauchy problem of the reduced field equations in harmonic
   coordinates.
\end{definition}
The smoothness of the quantities appearing in the definition is required,
since the transformation to harmonic coordinates involves derivatives.
\begin{proposition}\label{prop.adm}\ein
   Let $(\Sigma,h,k)$ be a symmetric initial data set, with matter obeying
   the strong energy condition and (compare equation \eqref{e.lambda} for
   definition) $\lambda>0$ somewhere on
   $\Sigma$. Further, let $t_0$ denote an arbitrary real number.\zeile
   Then there exists a $\delta>0$ and a 
   PMC foliated surface symmetric spacetime $(\bar{M},\bar{g})$
   diffeomorphic to  
   $]t_0-\delta,t_0+\delta[\,\times \Sigma$ with an embedding 
   $\iota : \Sigma \longrightarrow \bar{M}$, satisfying $\iota(\Sigma)=S_{t_0}$
   and $\iota_*h$, $\iota_*k$ are the first and second fundamental form of
   $S_{t_0}$ in $(\bar{M},\bar{g})$. $(\bar{M},\bar{g})$ obeys the strong
   energy condition and $\bar{g}$ can be written in  the form described in
   proposition \ref{prop.4geometry}. This construction is unique up to
   the choice of $t_0$ and $\delta$.
\end{proposition}
\begin{proof}\zeile
   On the induced manifold $\hat{\Sigma}$ diffeomorphic to 
   $S^1 \times \hat{F}$ the induced data is invariant under the group action,
   hence the Cauchy developments also and we get a surface symmetric
   Cauchy development of the data, admitting a symmetric local in time PMC
   foliation near $\Sigma$ on some time interval
   $]t_0-\delta,t_0+\delta[$ and allowing a set of coordinates stated in
   proposition \ref{prop.4geometry}. The uniqueness property stated above
   follows from the geometric uniqueness of solutions of Einstein's
   equations associated with the uniqueness of the PMC foliation, once the
   remaining degree of freedom has been fixed by the requirement
   $\iota(\Sigma)=S_{t_0}$. Note, that lapse is fixed by the ellipticity of
   the respective PMC equation and shift is fixed by equation \eqref{e.s}
   and the condition $\nu(t,0) = 0$.
\end{proof}

In surface symmetric spacetimes, there is another way to express the
constraint equations \eqref{e.ch} and \eqref{e.cm} in terms of 'optical
scalars', see \cite{gm} for an enlightening presentation.\zeile 
The geodesic null congruences determined by $k_{\pm}=m \pm n$
give rise to the null expansions
\begin{equation}\label{e.expansions}
   \vartheta_{\pm} := - (\tr\lambda \pm \tr\kappa)
                    = \frac{2}{r} (m(r) \pm n(r))
                    = \frac{2}{r} k_{\pm}(r)
                    = 2 A^{-2}A' \mp (H-K)
   \komma
\end{equation}
where \eqref{e.lambdakappa}, with \eqref{e.trkappa} has been used, together
with the relation $r=Aa$ for the area radius defined in
\eqref{e.radius}. The formula illustrates the definition of $r$ as a volume
measure, whose variation along $k_{\pm}$ is descibed by the
(negative of the) trace of the second fundamental form associated to this
direction.\zeile
Now we can write the constraint equations 
for an arbitrary symmetric Cauchy surface $S$ symmetrically as
\begin{gather*}
   8\pi(\rho+j) = - m(\vartheta_-) - \tfrac{3}{4}\vartheta_-^2
                  + \vartheta_- H + r^{-2}\epsilon \\
   8\pi(\rho-j) = - m(\vartheta_+) - \tfrac{3}{4}\vartheta_+^2
                  - \vartheta_+ H + r^{-2}\epsilon
\end{gather*}
and taking $\omega_{\pm}:=r\vartheta_{\pm}$
as the fundamental variable we get (compare \cite{gm})
\begin{equation}\label{e.momega}
   m(\omega_{\pm}) = -8\pi r(\rho \mp j) \mp \omega_{\pm}H +
      \tfrac{1}{4r}
      \left( \omega_+\omega_- - 2\omega_{\pm}^2 + 4\epsilon \right)
   \punkt
\end{equation}

Furthermore, the area radius serves as a warping function in the warped
product $M = B \times_r F$ of the two dimensional spacetime $(B,\ubar{g})$
with $(F,\obar{g})$, where
$B=\RR \times S^1$ is the quotient $M/G$ and $\ubar{g}=g_{|B}$. We adopt
the convention that lower case Latin letters from the beginning of 
the alphabet range from 0 to 1, and objects intrinsic or orthogonally
projected to $(B,\ubar{g})$ 
will be marked by an underbar. Einsteins equations in this framework can be
considered as equations in $(B,\ubar{g})$ for the field 
$r: B \longrightarrow \RR$:
\begin{equation}\label{e.fieldr}
   \ubar{\nabla}_a \ubar{\nabla}_b\, r 
      = \frac{\mass}{r^2} \ubar{g}_{ab}
        - 4\pi r \left( \ubar{T}_{ab} - \tr \ubar{T} \ubar{g}_{ab} \right)
   \komma
\end{equation}
where $\ubar{T}$ denotes the projected energy-momentum tensor into the
spacetime $(B,\ubar{g})$ and the mass function \mass is defined as 
\begin{equation}
   \label{e.mass}
   \mass := \frac{1}{2}r \left(\epsilon - \ubar{\nabla}^a r 
                                      \ubar{\nabla}_a r \right)         
      = \frac{1}{2}r \left(\epsilon - \viernabla^{\alpha} r\,
                                      \viernabla_{\alpha} r \right) 
      = \frac{1}{2}r 
           \left(\epsilon - \tfrac{1}{4} r\vartheta_+ r\vartheta_- \right) 
   \komma        
\end{equation}
since 
$\vartheta_+\vartheta_- = 
  4/r^2 \,\viernabla^{\alpha} r \viernabla_{\alpha} r$.
$\mass$ turns out to be the Hawking mass $m_H(F)$ (up to a factor
$-\frac{1}{2}\chi$ for $\text{genus}(F) \ge 3$):
\begin{equation}\label{e.hawkingmass}
   m_H(F) := \frac{\vol(F)^{1/2}}{(4\pi)^{3/2}} 
             \left( \pi\chi(F) - 1/8\int_F \vartheta_+\vartheta_- \right)
\end{equation}
and differentiation of \eqref{e.mass} yields the mass flux equation
\begin{equation}
   \label{e.massflux}
   \ubar{\nabla}_a \mass 
      = 4\pi r^2 \left( \ubar{T}_{ab} - \tr \ubar{T} \ubar{g}_{ab} \right) 
                 \ubar{\nabla}^b r
   \punkt
\end{equation}
\subsubsection{Expanding and recollapsing models}\label{ss.expanding}
Let us first consider the definition of mass in \eqref{e.mass}. In the
spherically symmetric case we see, that $\grad r$ is spacelike as long as 
$2\mass<r$ holds, a condition we are familiar with in connection with the
Schwarzschild spacetime. 

In the plane and hyperbolic case the picture is quite different, since
$\grad r$ 
turns out to be timelike, as long as $2\mass/r>\epsilon$. In fact, the
lemmas 2.3, 2.4 and 2.5 in \cite{r4} prove for spacetimes with hyperbolic
symmetry and in non-flat plane symmetric spacetimes $\grad r$ is timelike,
provided the dominant energy condition is fulfilled. Thus we cannot think
about $r$ as some radial, spacelike coordinate any longer and this fact will
play a central role in our further analysis:\zeile
We can choose the time orientation in those spacetimes to find
$\grad r$ past pointing, {\em throughout the whole spacetime}. Then we
define the time orientation on the cotangent bundle by 
metric transport from the tangent bundle, such that $dr$ turns out to be
future pointing. Therefore $r$ increases with time, which means, that the
area of the surfaces of symmetry increases with time, and the spacetime
expands in this sense. Thus we will call these spacetimes expanding. 
We get an equivalent characterisation by the relation between $\grad r$  
and the null expansions: $\vartheta_+$ and $\vartheta_-$ have
fixed and opposite signs (in particular $\vartheta_+>0$, $\vartheta_-<0$),
since $\vartheta_{\pm}=(2/r)k_{\pm}(r)=(2/r)dr(k_{\pm})$ is the contraction
of a timelike with a null vector.\zeile
Therefore we can decide, given a symmetric initial data set $(\Sigma,h,k)$
with non-flat plane or hyperbolic symmetry, with matter obeying the
dominant energy condition, which direction is expanding or contracting.
This is important, since we expect some singularity towards the contracting
direction and therefore we will pay attention to the past
development $D^-(\Sigma)$ which represents the contracting direction with 
respect to our conventions.

Note, that we obtained this information without referring to the
mean curvature $H=\tr k$. But it turns out, that the mean curvature of
an arbitrary Cauchy surface $S$ in expanding spacetimes is also restricted
in some way:\zeile
First remember, that $H=\tr k=-\div n$ measures the convergence of the
geodesic congruence, future pointing and orthogonal to $S$. Thus, $H<0$
everywhere on $S$ corresponds to the notion of expansion and Hawking's
theorem proves the existence of a singularity in the past. In an
expanding spacetime in our sense, $H$ is not necessarily
confined to be negative everywhere. But 
we will see, that it is impossible for $H$ to become non-negative
everywhere on $S$: 
The explicit formula \eqref{e.expansions} for $\vartheta_{\pm}$ shows in
connection with our sign conventions, that $H<K$. If $H$ were non-negative,
then $\abs{H}<\abs{K}$, thus $H^2-K^2<0$ and integration of the Hamiltonian
constraint \eqref{e.ch} over $S^1$ yields a contradiction.\zeile
With the dominant energy condition we can show more. Writing the
Hamiltionian constraint as $R + H^2 -\abs{k}^2 = 16\pi\rho$ we get 
$R+H^2 \ge 0$. Assuming $H \equiv 0$ on $S$, we would get $R \ge 0$. 
But symmetric surfaces $S$ in spacetimes with plane and hyperbolic symmetry
obey the topological conditions (i), respectively (ii), of theorem 5.2. in
\cite{sy}, which imply, that $S$ cannot have positive scalar curvature and
must be flat in case of non-negative scalar curvature. Thus $S$ must be
flat and the scalar curvature vanishes. This in turn forces $\rho=0$ and
$k=0$ by the Hamiltonian constraint. In the plane symmetric case then the
spacetime is flat by $k=0$, $\rho=0$ and the dominant energy condition,
contradiction. In the hyperbolic symmetric case integration of equation
\eqref{e.ch} yields a contradiction, too.

Putting all this together we conclude without loss of generality, that surface
symmetric spacetimes, obeying the dominant energy condition, which are
plane symmetric and not flat or have hyperbolic symmetry are everywhere
expanding, with $dr$ timelike future pointing, 
$\vartheta_+>0$, $\vartheta_-<0$ and any symmetric Cauchy surface $S$ is
not maximal with mean curvature not everywhere positive on $S$.

Of course, these arguments do not work in the spherically symmetric case, 
where $\grad r$ is spacelike in $\{2\mass<r\}$ and no fixed sign of the
expansions $\vartheta_{\pm}$ can be expected, fitting into the the general
belief in the closed universe recollapse conjecture, which precludes 
expansion of the whole spacetime. In particular we expect the existence of
a maximal hypersurface in $M$.
%
%
%
%
%
%
%
%
%
%
%\pagebreak
\subsection{A priori estimates for the f\/ield equations}\label{s.estimates}
Our aim now is to get sufficient estimates, that allow the
construction of a global PMC foliation. The 'size' of the
foliation is measured by the mean curvature $H$, thus \emph{we are looking for
\emph{uniform} estimates of the geometric and matter 
quantities in terms of $H$}.   

So let $(M,g)$ be a surface symmetric spacetime, satisfying the dominant
and 
strong energy condition. Let us assume, that there is a Cauchy surface
$\Sigma$ in $M$ with mean curvature $H<0$. In particular we get a local in
time PMC 
foliation $\{S_t\}$, $t \in \,]t_1,t_2[$ with $\Sigma=S_0$ by theorem
\ref{thm.pmclocal}. If the 
spacetime possesses plane or hyperbolic symmetry, we choose the time
orientation in correspondence to the conventions introduced in
\ref{ss.expanding}, thus $H$ decreases with decreasing PMC time. If $(M,g)$ is
spherically symmetric we choose the time orientation, that $H$ decreases
with PMC time, too. Then in either case we expect to find a singularity at
least in the past $D^-(\Sigma)$ of $\Sigma$.

In $D^-(\Sigma)$ the mean curvature is bounded from above by 
$H \le \bar{H} < 0$ and $H=\bar{H}$ only on $\Sigma$. Thus $\abs{H}$ is
bounded from below and we find the following estimates for the field
equations in $D^-(\Sigma)$ as long as $H$ remains finite:
\begin{itemize}
\item At first we consider the constraint equation \eqref{e.momega} on
   a fixed leaf. At the critical points of $\omega_{\pm}$ we find together
   with the dominant energy condition the important inequality 
   \begin{equation}\label{e.mom}
      \abs{r\vartheta_{\pm}} \le 
      2\left( \abs{Hr} + \sqrt{(Hr)^2 + \epsilon} \right)
   \end{equation}
   as shown in \cite{r4}.

   For plane and hyperbolic symmetry this inequality \eqref{e.mom} can be
   strengthened to 
   \begin{equation*}
      \abs{\vartheta_{\pm}} \le 4\abs{H} \le C
      \quad \Longrightarrow \quad
      \abs{A^{-2}A'} \le C \,\text{ and }\, \abs{K} \le C
   \end{equation*}
   by the definition \eqref{e.expansions} of the null expansions.\zeile
   In the spherically symmetric case the argument is more complicated:
   First, the work \cite{b} of Burnett shows, that under the additional
   assumption of the non-negative-pressures condition we have $r \le C$ and
   $0 < C \le \mass$. Using the upper bound for $r$ on the right-hand side of
   \eqref{e.mom} we get with \eqref{e.mass}
   \begin{equation*}
      \abs{r\vartheta_{\pm}} \le C
      \quad \Longrightarrow \quad
      \frac{\mass}{r} \le C
   \end{equation*}
   Thus we get $r^{-1} \le C \mass^{-1} \le C$, hence
   $r$ is bounded above and below away from zero. Inserting this in the
   estimate \eqref{e.mom} leads to
   $\abs{\vartheta_{\pm}} \le C r^{-1} \le C$ and we are in the same
   situation as in the plane/hyperbolic case, hence $\abs{A^{-2}A'} \le C$
   and $\abs{K} \le C$. 
\item Now consider the lapse equation \eqref{e.l} on a fixed leaf. At
   the point, where $N$ attains its maximum $\bar{N}$, we have 
   \begin{equation*}
      \bar{N} \le \Big( \tfrac{1}{2}(H-K)^2 + K^2
                        + 4\pi(\rho+\tr S)
                  \Big)^{-1} 
              \le C/H^2 \le C
   \komma
   \end{equation*}
   due to the strong energy condition. Hence we have a bound for
   $\abs{N}$. 
\item Next, there is a bound for shift. Examination of formula \eqref{e.s}
   shows, that all quantities on the right hand side are bounded, so
   we get $\abs{\nu'} \le C$, and therefore, using $\nu(t,0)=0$: 
   $\abs{\nu(t,x} \le \abs{\nu(t,0)} + \int_{S^1} \abs{\nu'} \le C$.
\item By the way, the equation \eqref{e.p'} for $H'$ provides a bound for
   $\abs{H'}$, since the coefficient on the right-hand side is bounded
   and applying Gronwall gives the desired estimate.
\item With the information about $\nu'$ examination of \eqref{e.e2},
   $\dot{a}=a \left(-\tfrac{1}{2}N(H-3K) - \nu'\right)$, shows, that the
   factor in 
   brackets is already bounded and we get an inequality of the form
   $\abs{\d_t \ln \abs{a}} \le C$, which leads to a bound for $\abs{a}$
   and $\abs{a^{-1}}$.
\item The same line of argumentation works for
   $A$. Equation \eqref{e.e1} can be written as $\dot{A}-\nu A' = A (-NK +
   \nu')$, with the factor on the right-hand side bounded. So we
   get bounds for $\abs{A}$ and $\abs{A^{-1}}$, too.\zeile 
   Since we have already bounded the null expansions $\vartheta_{\pm} = 2
   A^{-2}A' \pm (H-K)$, one sees easily, that even $\abs{A'} \le C$.
\item Integration of equation \eqref{e.ch} over
   the circle yields an inequality 
   \begin{align*}
      \tfrac{1}{8}\int_{S^1} A^{5/2} 16\pi\rho 
        & = \tfrac{1}{8}\int_{S^1} A^{5/2}
            \Big(H^2-K^2-\tfrac{1}{2}(H-K)^2\Big) 
          + \tfrac{1}{4}\int_{S^1} A^{1/2}a^{-2}\epsilon \\
        & \le \tfrac{1}{8}\int_{S^1} A^{5/2}H^2 
            + \tfrac{1}{4}\int_{S^1}A^{1/2}a^{-2}
   \punkt
   \end{align*}
   From this one concludes the
   boundedness of $\int_{S^1}\rho$, and by the dominant energy condition
   the boundedness of $\int_{S^1} \abs{j}$ and $\int_{S^1}
   \abs{S}$. Integrating now 
   equation \eqref{e.l} starting at a point, where $N'=0$, we get 
   \begin{align*}
      N' & = -\int A^{-1}A'N' 
             +\int A^2N \Big(\tfrac{1}{2}(H-K)^2 + K^2 
                              +4\pi(\rho+\tr S)\Big)
             -\int A^2 \\
      \abs{N'} & \le C + \int \abs{A^{-1}A'N'}
      \punkt
   \end{align*}
   The bound for $\abs{N'}$ then
   follows from Gronwall's inequality.
\item Furthermore, the bounds for $A$,$A^{-1}$ together with the
   basic estimate for $\vartheta_{\pm}$ and the boundedness of
   $\int_{S^1} \rho$ are enough 
   to apply the proof of the lemma in \cite{r5}, which ends up
   with $\abs{N^{-1}} \le C$.
\end{itemize}

Collecting all these estimates we get the
\begin{proposition}\label{prop.first_estimates}\ein
   Let $(M,g)$ be a surface symmetric spacetime, obeying the dominant and
   strong energy condition and in the spherically symmetric case the
   non-negative-pressures condition, too. Assume the existence of a
   symmetric Cauchy surface $\Sigma$ with strictly negative mean
   curvature. In particular we get from proposition
   \ref{prop.4geometry} a PMC time  
   coordinate $t$, ranging in $]t_1,t_2[$ with $\Sigma=\{t=0\}$ and $H$
   decreases with decreasing $t$.\zeile
   Then we have uniformly on $]t_1,0]$ 
   \begin{equation*}
      \abs{A},\abs{A^{-1}},\abs{A'},\abs{a},\abs{a^{-1}},
      \abs{H},\abs{H'},\abs{K},
      \abs{N},\abs{N^{-1}},\abs{N'},\abs{\nu},\abs{\nu'}
      \le C
      \punkt
   \end{equation*}
\end{proposition}

To put this result into some framework, we establish some
formalism, to have useful abbreviations at hand as well as to make clear
the dependence between estimates of geometric quantities and matter
variables. 

{\samepage
\begin{definition}\label{def.collect}\ein\nopagebreak\vspace{-0.4cm}
   \begin{equation*}
      \mathcal{F} := \left(
              A,a,N,\nu,H,K
           \right)
   \end{equation*}
   collects the quantities describing the geometry of the foliation
   and
   \begin{equation*}
      \Phi := \left(
                 \rho,j,S
              \right)
   \end{equation*}
   abbreviates the matter quantities.
\end{definition}
} % samepage

We have already estimated the quantities $\mathcal{F}$,
$A^{-1},a^{-1},N^{-1}$, as well as $A',N',\nu', H'$ and
$\dot{A},\dot{a},\dot{H}$ (by inspection of the field equations).
The idea is now, to bound all quantitities
$\mathcal{F},\Phi$ together with all of their derivatives 
uniformly on the time interval $]t_1,0]$. Then there exists a smooth
extension to the closure of the interval, which serves as a new symmetric
initial data set for the field equations in the sense described in
definition \ref{def.admdata}. Note, that the bounds for
$A^{-1},a^{-1},N,N^{-1}$ ensure, that the geometry remains regular at the
boundary of the time interval and the
$C^{\infty}-$bounds for lapse and shift turn out to be necessary to obtain
$C^{\infty}-$bounds for the fundamental forms.\zeile
Proposition \ref{prop.adm} then sets us in the position to extend the 
foliation at least in the past direction, where $H < 0$ holds:
Construct the solution stated in proposition \ref{prop.adm} and embed $M$
into the maximal Cauchy development of the data. 
 
To carry out this program, we need some knowledge about the matter
quantities. First, the matter has to obey the dominant and strong energy
conditions. Second, assume that the regularity of the geometry guarantees
the regularity of the matter in a certain way:
For all non-negative integers $m$ and $n$ we have 
\begin{equation}\label{e.matter_regularity}
   \begin{aligned}
      {}& \abs{\d_t^m \d_x^n {\cal F}} \le C 
      && \quad\Longrightarrow\quad
      \abs{\d_t^m \d_x^n \Phi} \le C \\
      {}& \abs{\d_t^m \d_x^{n+1} {\cal F}} \le C 
      && \quad\Longrightarrow\quad
      \abs{\d_t^{m+1} \d_x^n \Phi} \le C 
      \komma
   \end{aligned}
\end{equation}
then the following lemmas hold.
\begin{lemma}\label{lem.dx}\ein
   Assume, that the matter fulfills \eqref{e.matter_regularity}. Then for
   arbitrary non-negative integers $m,n$
   \begin{equation*}
      \begin{aligned}
         \forall_{k<m}\forall_l\; & \abs{\d_t^k \d_x^l {\cal F}} \le C \\
         \forall_{l \le n}\; & \abs{\d_t^m \d_x^l {\cal F}} \le C         
      \end{aligned}
      \quad\Longrightarrow\quad
      \abs{\d_t^m \d_x^{n+1} {\cal F}} \le C 
   \end{equation*}
   holds.
\end{lemma}
\begin{lemma}\label{lem.dt}\ein
   Assume, that the matter fulfills \eqref{e.matter_regularity}. Then for
   arbitrary non-negative 
   integers $m,n$
   \begin{equation*}
      \begin{aligned}
         \forall_{k<m}\forall_l\; & \abs{\d_t^k \d_x^l {\cal F}} \le C \\
         \forall_{l \le n+1}\; & \abs{\d_t^m \d_x^l {\cal F}} \le C 
      \end{aligned}
      \quad\Longrightarrow\quad
      \abs{\d_t^{m+1} \d_x^n {\cal F}} \le C 
   \end{equation*}
   holds.
\end{lemma}
Together with proposition \ref{prop.first_estimates} the lemmas
accomplish the 
task of bounding all necessary derivatives of the geometric
quantities, by first bounding all spatial derivatives $\d_x^n{\cal F}$
(proposition \ref{prop.first_estimates} and lemma \ref{lem.dx}) and
then successively all derivatives $\dd{k}{l}{\cal F}$, with $k+l=n$ by
alternative applications of the lemmas.
Therefore, in view of proposition \ref{prop.first_estimates} the validity of
property \eqref{e.matter_regularity} will be enough to extend the
foliation. 
We will prove \eqref{e.matter_regularity} for some
matter models in the next section. But first, of course, we have to prove
the lemmas.

\begin{proof}[Proof of lemma \ref{lem.dx}]\zeile\vspace*{-0.6cm}
   \begin{itemize}
   \item[A:] Applying $\dd{m}{n}$ on equation \eqref{e.ch} and
      integrating along $S^1$ yields a bound for the difference
      $\abs{ (\dd{m}{n}A^{1/2})'(y) - (\dd{m}{n}A^{1/2})'(x) }$ 
      and using $\int_{S^1} (\dd{m}{n}A^{1/2})' = 0$ (hence is bounded),
      gives a bound for $\dd{m}{n+1}A^{1/2}$, hence for 
      $\dd{m}{n+1}A$.
   \item[a:] trivial
   \item[$\nu$:] Apply $\dd{m}{n}$ on \eqref{e.s}, then the right-hand
      side of \eqref{e.s} is bounded by assumption, from which the claim
      follows immediately.
   \item[N:] Again applying $\dd{m}{n}$ on the lapse equation
      \eqref{e.l} yields an equation of the form \eqref{e.l} for
      $\dd{m}{n}N''$ plus some already bounded term (by assumption and
      the already proven bound for $\dd{m}{n}A'$). The boundedness of
      $\dd{m}{n}N'$ follows then from Gronwall's estimate after
      integrating the equation for $\dd{m}{n}N''$ along $S^1$ starting at a
      point with $\dd{m}{n}N'=0$.
   \item[H:] Differentiation of equation \eqref{e.p'} gives an expression
      of the form 
      $(\d_t-\nu\d_x)\dd{m}{n}H'=B_1\dd{m}{n}H'+B_2$, with $B_1,B_2$
      bounded, where the assumptions and the already obtained bound for 
      $\dd{m}{n+1}\nu$ have been used. Thus applying Gronwall's inequality
      yields a bound for $\dd{k}{n+1}H$.
   \item[K:] Apply $\dd{m}{n}$ on equation \eqref{e.cm}, then the
      right-hand side is bounded by the previous estimates, hence
      bounding $\dd{m}{n+1}K$.
   \end{itemize}\vspace*{-0.6cm}
\end{proof}

\begin{proof}[Proof of lemma \ref{lem.dt}]\zeile\vspace*{-0.6cm}
   \begin{itemize}
   \item[A:] Applying $\dd{m}{n}$ on \eqref{e.e1} yields immediately a
      bound for $\dd{m}{n}\dot{A}$, since the right-hand side is already
      bounded.
   \item[a:] The same argument with $\d_t^m$ instead of $\dd{m}{n}$
      applied to equation \eqref{e.e2} works in this case to bound
      $\dd{m}{n}\dot{a}$. 
   \item[H:] And again, $\dd{m}{n}$ on equation \eqref{e.p}, yields an
      estimate for $\dd{m}{n}\dot{H}$.
   \item[K:] Apply $\dd{m}{n}$ on equation \eqref{e.e3}, then the only
      terms on the right-hand side, not already known to be bounded are
      $\dd{m}{n}N''$ and $\dd{m}{n}A''$. But inserting equations
      \eqref{e.l}, respectively \eqref{e.ch} for $N''$ and $A''$, one
      sees easily, that $\dd{m}{n}N''$ and $\dd{m}{n}A''$ indeed are
      bounded by the right-hand sides of their equations, hence
      $\dd{m+1}{n} K$ is bounded.
   \item[N:] Unfortunately there is no
      explicit equation for $\dot{N}$. So we have to apply the derivative
      operator $\dd{m+1}{n}$ to equation \eqref{e.l}, therefore
      producing already estimated terms involving $\dd{m+1}{n}$
      applied on the quantities treated above, as well as on $\Phi$
      (bounded by property \eqref{e.matter_regularity}), but 
      also a term involving $\dd{m+1}{n+1}A$ 
      on the right-hand side. If this term turns out to be bounded,
      then the equation for $\dd{m+1}{n}N''$ has the form 
      \begin{align*}
         \dd{m+1}{n} N'' & = B -A^{-1}A' \dd{m+1}{n} N' \\
         & + A^2 \dd{m+1}{n} N 
         \Big( \tfrac{1}{2}(H-K)^2 + K^2
               + 4\pi(\rho+\tr S) 
         \Big) 
         -A^2
%         \komma
      \end{align*}
      with $\abs{B}$ bounded. On $S^1$, $\dd{m+1}{n} N$ attains
      its maximum, from which we can infer
      \begin{equation*}
         \dd{m+1}{n} N \le \left\{ (1-B/A^2)
           \Big( \tfrac{1}{2}(H-K)^2 + K^2
                 + 4\pi(\rho+\tr S) 
           \Big)^{-1}\right\}_{\max}
%        \komma
      \end{equation*}
      hence it is bounded from above. Similarily for the minimum:
      \begin{equation*}
         \dd{m+1}{n} N \ge \left\{ (1-B/A^2)
           \Big( \tfrac{1}{2}(H-K)^2 + K^2
                 + 4\pi(\rho+\tr S) 
           \Big)^{-1}\right\}_{\min}
%        \komma
      \end{equation*}
      which is bounded from below and we get $\abs{\dd{m+1}{n} N} \le
      C$.\zeile 
      It remains to show the boundedness of 
      $\dd{m+1}{n+1}A = \dd{m}{n+1}\dot{A}$. Inserting equation 
      \eqref{e.e1} for $\dot{A}$, we see on the right-hand side beside a
      bounded term the quantities $\dd{m}{n}A''$ and
      $\dd{m}{n+1}\nu'$. Inserting equation \eqref{e.ch} for $A''$ and
      equation \eqref{e.s} for $\nu'$, we finally obtain bounds for
      $\dd{m}{n}A''$ and $\dd{m}{n+1}\nu'$, hence for
      $\dd{m}{n+1}\dot{A}$, as desired.
   \item[$\nu$:] The final estimate is straightforward: Apply the
      operator $\dd{m+1}{n-1}$ to equation \eqref{e.s}, which yields
      immdediately a bound for $\dd{m+1}{n}\nu$ since the right-hand
      side of the equation is already bounded by the arguments above.
   \end{itemize}\vspace*{-0.6cm} 
\end{proof}
%
%
%
%
%
%
%
%
%
%
%\pagebreak
\subsection{Higher order estimates}
Here we prove the matter regularity condition
\eqref{e.matter_regularity} for Einstein-Vlasov, Einstein-Maxwell and
the Einstein-Vlasov-Maxwell system, achieving the goal of this work. 
\subsubsection{Collisionless matter}\label{s.vlasov}
The 3+1-split of the Vlasov equation \eqref{e.vlasov} with respect to
the symmetries reads 
\begin{equation}\label{e.surfvlasov}
      \begin{aligned}
         0 & = \d_t f + \left( NA^{-1}\tfrac{v^1}{v^0}-\nu \right)\d_x f\\
              & + \left( -A^{-1}N'v^0 + NKv^1 + NA^{-2}A'
                     \tfrac{(v^2)^2+(v^3)^3}{v^0}
                  \right) \frac{\d}{\d v^1} f\\
              & - N \left( A^{-2}A'\tfrac{v^1}{v^0} - \tfrac{1}{2}(H-K)
                    \right)v^B \frac{\d}{\d v^B} f
         \komma
      \end{aligned}
   \end{equation}
where $v^0 := \sqrt{1+\delta_{ij}v^iv^j}$ and 
$0 = \d_{x^2}f = \d_{x^3}f = v^3\tfrac{\d}{\d v^2}f - v^2\tfrac{\d}{\d v^3}f$
(by symmetry) has been used.\zeile
The energy momentum tensor \eqref{e.vlasovenergy} associated with the
Vlasov equation leads to the matter quantities
\begin{gather*}
   \rho = \int fv^0\,dv \\
   j = \int fv^1\,dv \\
   S_{ab} = \int fv_av_b/v^0\,dv
   \punkt
\end{gather*}

Now we investigate the matter regularity property
\eqref{e.matter_regularity}. Since $f$ is constant along the
characteristics of the Vlasov equation, given an initial particle 
distribution $f_0$ on the mass shell over $\Sigma$, we get the matter
distribution for each time $t$ as a function $f(t,y,w)$ on the mass
shell over the leaf $S_t$, by 
$f(t,y,w) = f_0(Y(t,y,w),V(t,y,w))$, where $(Y,V)$ denote the
characteristic curve  
of the Vlasov equation through the point $(0,y,w) \in P_{(0,y)}$,
$(0,y) \in \Sigma$.\zeile
Thus the matter quantities $\rho,j,S$ are bounded, provided the
support of $f$ remains bounded. Define now 
\begin{equation*}
   \bar{P}_f(t):=\{\sup\abs{v}\,|\, v \in \supp f(t,y,w)\;
   \forall_{(0,y,w) \in P_{(0,y)}} \}
   \komma   
\end{equation*}
then the matter quantities are bounded by $C(1+\bar{P}_f(t))^4$, hence
it is sufficient, to control $\bar{P}_f(t)$ on $]t_1,0]$.\zeile
Since the characteristic curves $(Y,V)$ are integral curves of $X$ and 
all coefficients in the components of $X$ are already bounded by
proposition \ref{prop.first_estimates},
the characteristics itself are bounded, $\bar{P}_f(t) \le C$ as desired.

Now we need to iterate this procedure. Assume, that all 
$\dd{k}{l}{\cal F}$ and $\dd{k}{l}\Phi$ for $k+l=m+n$ are bounded. 
To bound the derivatives of the matter quantities of order $m+n+1$, 
differentiate the Vlasov equation $m+n+1$ times (with respect to $t$ and $x$
only). This yields linear equations for $\dd{\bar{k}}{\bar{l}}f$, for all
non-negative integers $\bar{k}+\bar{l}=m+n+1$, of the form
\begin{equation*}
   X (\dd{\bar{k}}{\bar{l}} f) = B
   \komma
\end{equation*}
where B vanishes outside the support of $f$, which bounds the support of
$B$. Moreover, $B$ involves derivatives of order $m+n+1$ of the quantities 
${\cal F}$ and $A'$, $N'$. For derivatives of order at most $m$ in $t$, we
see the boundedness of $\abs{B}$, by applying similar arguments as used in
the proof of lemma \ref{lem.dx}. Thus in this case
$\dd{\bar{k}}{\bar{l}}f$ is 
bounded and $\bar{P}_{\dd{\bar{k}}{\bar{l}} f}$, since they have the same
characteristics, hence the derivatives of the matter quantities of order
$m+n+1$, involving derivatives of order at most $m$ in $t$ are
bounded. Moreover, due to the simple dependence of the characteristics on
the frame variables $v$, all derivatives of $f$ with respect to $v$ are
bounded, too.\zeile
In view of this fact, we are able to bound all derivatives of order $m+n+1$
of the matter quantities, by considering the Vlasov equation as an equation
for $\dd{m+1}{n+1} f$, for which the right-hand side is known to be
bounded.\zeile 
Therefore the first part of property \eqref{e.matter_regularity} holds.
For the second part we must show, that the spacelike derivatives of 
${\cal F}$ can be redistributed to timelike derivatives of $\Phi$. But this
has already been done in the proof of the first part of
\eqref{e.matter_regularity}, and we get the 
\begin{theorem} \label{thm.ev} \ein
   Let $(M,g,f)$ be a surface symmetric solution of the
   Einstein-Vlasov system, which
   possesses a symmetric Cauchy surface $\Sigma$ with strictly
   negative mean 
   curvature $H \le \bar{H} < 0$ and $H=\bar{H}$ somewhere on
   $\Sigma$.\zeile 
   Then all of the past of $\Sigma$ admits a PMC foliation
   $\{S_t\}$, where $t$ takes all values in the interval
   $\,]\!-\infty,0]$ and $H$ takes all values in
   $\,]\!-\infty,\bar{H}]$.
\end{theorem}
\subsubsection{Maxwell field}
Let $F$ denote the electromagnetic field. The symmetry simplifies $F$:
Due to the symmetries $F$ can be written relative to the orthonormal
coframe $\{\viersigma^{\mu}\}$ introduced in section \ref{s.geometry} as
\begin{equation*}
   F = -\hat{e}(t,x)\, \viersigma^0 \wedge \viersigma^1
       +\hat{b}(t,x)\, \viersigma^2 \wedge \viersigma^3
   \komma
\end{equation*}
since all other components are forced to vanish and $F$ and $\viersigma^0
\wedge \viersigma^1$, $\viersigma^2 \wedge \viersigma^3$ 
remain invariant under the action of the symmetry group.\zeile
To obtain an explicit form of the Maxwell equations $F$ and $*F$ are
represented in the given coordinates by
\begin{align*}
&
\begin{gathered}
   F_{\alpha\beta} \sim
     \begin{pmatrix}
        0 & \vline & -e &  0 & 0 \\ \hline 
        e & \vline &  0 &  0 & 0 \\  
        0 & \vline &  0 &  0 & b \\ 
        0 & \vline &  0 & -b & 0 
     \end{pmatrix}
\end{gathered}
&
\begin{gathered}
   e(t,x) = NA\, \hat{e} \\
   b(t,x,\vartheta) = (Aa)^2 \epsilon_{\vartheta}\, \hat{b}
\end{gathered} 
\\
&
\begin{gathered}
   *F_{\alpha\beta} \sim
     \begin{pmatrix}
        0         & \vline & \kappa b &  0             & 0 \\ \hline 
        -\kappa b & \vline &  0       &  0             & 0 \\  
        0         & \vline &  0       &  0             & \kappa^{-1} e \\ 
        0         & \vline &  0       & -\kappa^{-1} e & 0 
     \end{pmatrix}
\end{gathered}
&
\begin{gathered}
   \sqrt{\abs{g}} = NA^3a^2\abs{\epsilon_{\vartheta}} \\
   \kappa := NA^{-1}a^{-2}\abs{\epsilon_{\vartheta}^{-1}}
\end{gathered}
\punkt
\end{align*}

The Maxwell equations \eqref{e.maxwell} in vacuum are given by setting 
$J^{\alpha}=0$. We get for the magnetic field
\begin{equation*}
   \begin{aligned}
      F_{23}' & = 0 \\
      \dot{F}_{23} & = 0
   \end{aligned}
   \iff
   \begin{aligned}
      {} & b' & = 0 \\
      {} & \dot{b} & = 0
   \end{aligned}
   \punkt
\end{equation*}
Since $\epsilon_{\vartheta}$ does not depend on $(t,x)$, we find 
$\d_{\alpha}((Aa)^2\,\hat{b})=0$ for $\alpha=0,1$, hence
\begin{align*}
   \hat{b} & = C (Aa)^{-2} \\
   b       & = C \epsilon_{\vartheta}
   \punkt
\end{align*}
For the electric field we find
\begin{equation*}
   \begin{aligned}
      *F_{23}' & = 0 \\
      *\dot{F}_{23} & = 0
   \end{aligned}
   \qquad\Longrightarrow\qquad
   \begin{aligned}
      {} & (\kappa^{-1}e)' & = 0 \\
      {} & (\kappa^{-1}e)\dot{} & = 0
   \end{aligned}
   \punkt
\end{equation*}
(Note, that $*F_{01,2}$ vanishes, since $\kappa b$ contains no
$\epsilon_{\vartheta}$).
Again, since $\epsilon_{\vartheta}$ does not depend on $(t,x)$, we get
$\d_{\alpha}((Aa)^2\,\hat{e})=0$, for $\alpha=0,1$,  hence
\begin{align*}
   \hat{e} & = C\, (Aa)^{-2} \\
   e       & = C\, NA^{-1} a^{-2}
   \punkt
\end{align*}
Since $N,A^{-1},a^{-1}$ are already bounded, the same is true for the
electromagnetic fields $e$ and $b$.

The energy momentum tensor \eqref{e.maxwellenergy} takes the simple form
\begin{equation*}
   E(\viere_{\alpha},\viere_{\beta}) = \frac{1}{8\pi}(\hat{e}^2+\hat{b}^2)
      \begin{pmatrix}
         1 & 0 & 0 & 0 \\
         0 &-1 & 0 & 0 \\
         0 & 0 & 1 & 0 \\
         0 & 0 & 0 & 1 
      \end{pmatrix}
   \komma
\end{equation*}
and we get the matter quantities as
\begin{gather*}
   \rho = E(\viere_0,\viere_0) = -S_{11} = \tr S 
        = \frac{1}{8\pi}(\hat{e}^2+\hat{b}^2)
        = \frac{1}{8\pi}\left( C(Aa)^{-4} + C(Aa)^{-4}  \right) \\
   j = 0
   \punkt
\end{gather*}
The property \eqref{e.matter_regularity} posed in section \ref{s.estimates}
is obviously fulfilled, hence we get the
\begin{theorem} \label{thm.em}  \ein
   Let $(M,g,F)$ be a surface symmetric solution of the
   Einstein-Maxwell system with plane or hyperbolic symmetry, which
   possesses a symmetric Cauchy surface $\Sigma$ with strictly
   negative mean 
   curvature $H \le \bar{H} < 0$ and $H=\bar{H}$ somewhere on
   $\Sigma$.\zeile 
   Then all of the past of $\Sigma$ admits a PMC foliation
   $\{S_t\}$, where $t$ takes all values in the interval
   $\,]\!-\infty,0]$ and $H$ takes all values in
   $\,]\!-\infty,\bar{H}]$.
\end{theorem}
Note that the restriction to plane and hyperbolic symmetry is necessary,
since the electromagnetic energy momentum tensor does not obey the
non-negative 
pressures condition, and therefore fails to satisfy the assumptions of
proposition \ref{prop.first_estimates} in the case of spherical symmetry.
\subsubsection{Charged particles}
Now consider the Vlasov-Maxwell system in $(M,g)$. We obtain the coupled
equations by modifying the uncharged Vlasov equation \eqref{e.surfvlasov}
by adding the term 
\begin{align*}
   e \left( A^{-1} + A^{-2}A' 
           - N^{-1}\nu(1+A^{-1}A') \tfrac{v^1}{v^0} 
     \right) \frac{\d f}{\d v^1} 
\end{align*}
on the right-hand side. The matter current $J$
according to the energy-momentum tensor
\begin{equation*}
   \Theta = T + E
   \komma\qquad
   \begin{aligned}
      \rho & = & \Theta(\viere_0,\viere_0) \\
      j    & = & -\Theta(\viere_0,\viere_1)
   \end{aligned}
\end{equation*}
is of the form
\begin{equation*}
   J = \rho \viere_0 + j \viere_1 
     = N^{-1}\rho \d_0 + (A^{-1}j - N^{-1}\nu\rho)\d_1
   \punkt
\end{equation*}

The homogeneous Maxwell equations remain unchanged, which yields as before
\begin{align*}
   \hat{b} & = C (Aa)^{-2} \\
   b       & = C \epsilon_{\vartheta}
   \punkt
\end{align*}
Since 
$*J_{\alpha\beta\gamma} = 
   \sqrt{\abs{g}}\epsilon_{\mu\alpha\beta\gamma} J^{\mu}$, 
and again for $\alpha=0,1$,  
$\d_{\alpha}(\kappa^{-1}e)
 =\abs{\epsilon_{\vartheta}}\d_{\alpha}((Aa)^2\hat{e})$
the inhomogeneous equations read
\begin{equation*}
   \begin{aligned}
      {} & (\kappa^{-1}e)'      & = &&  4\pi\sqrt{\abs{g}}J^0 \\
      {} & (\kappa^{-1}e)\dot{} & = && -4\pi\sqrt{\abs{g}}J^1
   \end{aligned}
   \quad\Longleftrightarrow\quad
   \begin{aligned}
      {} & \hat{e}' & = & -2A^{-1}A'\,\hat{e} && + 4\pi A\rho \\
      {} & \dot{\hat{e}} & = & -2(Aa)^{-1}(Aa)\dot{}\,\hat{e} 
                               && + 4\pi (A\nu\rho - N j)
   \end{aligned}
   \punkt
\end{equation*}

To get a bound for $\hat{e}$, note that the inhomogeneous Maxwell equations
are of the form
\begin{equation*}
   \d_{\alpha}\hat{e} = \varphi_{\alpha}\,\hat{e} 
                       + \psi_{\alpha}\cdot\tbinom{\rho}{j}
   \komma\qquad
   \alpha=0,1
\end{equation*}
with $\varphi_{\alpha},\psi_{\alpha}$ bounded by proposition
\ref{prop.first_estimates}. Integrating the second equation gives
\begin{equation*}
      \left|\int_{S^1} \dot{\hat{e}}\right|
      = \left|\int_{S^1} \varphi_0\,\hat{e} 
        + \int_{S^1}\psi_0\cdot\tbinom{\rho}{j}\right|
      \le C \left|\int_{S^1} \hat{e}\right| + C 
   \komma
\end{equation*}
since $\int_{S^1} \rho, \int_{S^1} \abs{j} \le C$ by integrating equation
\eqref{e.ch} as shown in the proof of proposition \ref{prop.first_estimates}.
Using then the inequality
\begin{equation*}
   \left|\int_{S^1} \hat{e}\right|(t)
   \le \left|\int_{S^1} \hat{e}\right|(0)
     + \int_0^t \left|\int_{S^1} \dot{\hat{e}}\right|
\end{equation*}
and applying Gronwall's inequality on $\left|\int_{S^1} \hat{e}\right|$ we
obtain 
\begin{equation*}
   \left| \int_{S^1} \hat{e} \right| \le C
   \punkt
\end{equation*}
Now using the first equation, we get analogously 
\begin{equation*}
   \abs{\hat{e}(t,y)-\hat{e}(t,x)}
      = \left|\int_x^y \hat{e}'\right|
      = \left|\int_{S^1} \varphi_1\,\hat{e} 
        + \int_{S^1}\psi_1\cdot\tbinom{\rho}{j}\right|
      \le C \left|\int_{S^1} \hat{e}\right| + C
      \le C
   \komma
\end{equation*}
where $\left|\int_{S^1} \hat{e}\right| \le C$ has been used.

Thus we have
\begin{equation*}
   \begin{gathered}
      \left|\hat{e}(t,y)-\hat{e}(t,x)\right| \le C \\
      \left|\int_{S^1} \hat{e}\right| \le C      
   \end{gathered}
   \qquad\Longrightarrow\qquad
   \abs{\hat{e}} \le C
\end{equation*}
as desired.

Now property \eqref{e.matter_regularity} is obtained by bounding appropriate
derivatives of 
$\hat{e}$: Application of $\dd{m}{n-1}$ on the equation for $\hat{e}'$
shows the first part of property \eqref{e.matter_regularity} and
application of $\dd{m}{n}$ on the 
equation for $\dot{\hat{e}}$ establishes the second part. This proves:
\begin{theorem} \label{thm.evm}  \ein
   Let $(M,g,f,F)$ be a surface symmetric solution of the
   Einstein-Vlasov-Maxwell system with plane or hyperbolic symmetry, which
   possesses a symmetric Cauchy surface $\Sigma$ with strictly
   negative mean 
   curvature $H \le \bar{H} < 0$ and $H=\bar{H}$ somewhere on
   $\Sigma$.\zeile 
   Then all of the past of $\Sigma$ admits a PMC foliation
   $\{S_t\}$, where $t$ takes all values in the interval
   $\,]\!-\infty,0]$ and $H$ takes all values in
   $\,]\!-\infty,\bar{H}]$.
\end{theorem}
%
%
%
%
%
%
%
%
%
%
%\pagebreak
\subsection{Improving the results}\label{s.improve}
The purpose of this section is to get rid of the requirement of strictly
negative mean curvature on the Cauchy surfaces, which seems to be a rather
technical restriction. 
Let us instead assume $\Sigma$ to be a symmetric Cauchy surface in the
surface symmetric spacetime $(M,g)$ with
matter obeying the dominant and strong energy condition as well as the
non-negative pressures condition, such as the surface symmetric
Einstein-Vlasov system $(M,g,f)$.
Note, that we assume the non-negative pressures condition not only for
the spherically symmetric case. Indeed we will see, that this energy
condition is necessary for the arguments given in this section, thus
we cannot apply them to matter involving electromagnetic fields.\zeile 
As pointed out in \ref{ss.expanding} there is an important difference 
between the possible types of surface symmetry and it turns out to be a
good idea, to analyse the spherically symmetric case separately.
\subsubsection{Spherically symmetric spacetimes} \label{ss.spherically}
There are some results already obtained elsewhere, so we can exclude some
special cases from our analysis. Namely for the Einstein-Vlasov system (as
well as for the massless scalar field) it has been proven in \cite{r4} and
\cite{br}, that given an arbitrary symmetric constant mean curvature Cauchy
surface, there exist a global CMC foliation with the mean curvature taking
all real values. So we may assume, that the mean curvature $H$ on $\Sigma$
is not constant, and we immediately get from theorem \ref{thm.pmclocal} the
existence of a local in time PMC foliation $]t_1,t_2[\,\times \Sigma$ of a
neighbourhood of $\Sigma$ in $M$. 

There are some a priori estimates:
\begin{itemize}
\item It is shown in \cite{b}, that $r$
   and $\mass^{-1}$ are bounded. Theorem 2.1 in \cite{b} then shows, that all
   timelike curves have finite length.
\item Inserting the bound for $r$ on the right-hand side of the estimate
   \eqref{e.mom}, we get from \eqref{e.mass} 
   bounds for $r^{-1}$ and $\mass$ on any finite time interval.
\item Section III in \cite{br} shows, that in light of the fact, that all
   timelike curves have finite lengths, the volumes of two arbitrary
   Cauchy surfaces $\Sigma_1$ and $\Sigma_2$ are related by
   \begin{equation*}
        \vol(\Sigma_2) \le \vol(\Sigma_1)
           \left( 1 + C \sup_{\Sigma_1}\abs{H} \right)^3
        \punkt
   \end{equation*} 
   This implies, that
   the volumes of all Cauchy surfaces $S$ in $M$ are bounded above by the
   volume of $\Sigma$ and, since $H$ is bounded on each finite time
   interval, interchanging the roles of $\Sigma$ and $S$ shows, that
   the volumes are bounded from below, too, on each finite time interval.
\end{itemize}

If we denote the volume form of a PMC leaf $S_t$ by $\Omega$ and the volume
of $S_t$ by $V(t)$ we have $\Omega=\sqrt{\abs{h}}$ and
$V(t) = \int \Omega \,dx\,d\vartheta\,d\varphi
      = 4\pi a^{-1} \int_{S^1} r^3$.\zeile
With the bounds of $V,V^{-1},r,r^{-1}$ we get now $a,a^{-1} \le C$, and
again using the bounds for the radius function we get $A,A^{-1} \le C$,
thus the first fundamental form of the leaves is bounded from above and
below on each finite time interval.\zeile
Moreover, as shown in section \ref{s.estimates}, the estimate
\eqref{e.mom} bounds $\abs{A^{-2}A'}$ and $\abs{K}$, so that the
second fundamental form is also bounded (from above) on each finite time
interval as well as $\abs{A'} = \abs{A^2}\,\abs{A^{-2}A'}$.

Consider now the lapse equation \eqref{e.l}, written in the form
\begin{equation*}
   (AN')' = A^3 N 
      \Big( \tfrac{1}{2}(H-K)^2 + K^2
             + 4\pi(\rho+\tr S) 
      \Big) - A^3
   \komma
\end{equation*}
where the term in brackets is non-negative. Therefore, setting this term to
zero, we get the estimate $(AN')' \ge -C$, hence, for arbitrary 
$p,q \in S_t$: $(AN')(p)-(AN')(q) \ge -C$. Now choosing $q$ to be a
critical point of $N$, yields $N'(p) \ge -C$, and similarily choosing $p$
as a critical point of $N$ gives $N'(q) \le C$. From this we see, that
$\abs{N'}$ is bounded, since $p$, respectively $q$ is arbitrary on $S_t$.

The difference to the situation considered in section \ref{s.estimates} is,
that we are no longer confined to a mean curvature, having everywhere fixed
sign. The difficulty with the estimates done there
is, that they provide no information about the behaviour of the lapse
function, when $H$ becomes zero. The estimates 
done here so far, do not rely on this fact and the uniform bound of $N'$
shows, that either lapse remains finite or diverges uniformly to infinity.

In order to prove global existence of a symmetric PMC foliation assume,
that $]t_1,t_2[$ is the maximal time interval of existence. Without loss of
generality let us consider only a possible extension towards the
past, where $H$ decreases with decreasing $t$. Thus, we are looking for regular
symmetric initial data for $t=t_1$ in the sense of 
definition \ref{def.admdata}.\zeile 
Then there are two cases: First, $H(t)$ is everywhere positive or
everywhere negative near $t_1$, the
arguments in section \ref{s.estimates} apply (the fixed sign of $H$ is
enough to perform the estimates, whether $H$ is positive or negative),
extending the 
foliation and we get a contradiction to the maximality of the time interval,
hence $t_1=-\infty$.\zeile 
Second, $\lim_{t \rightarrow t_1} H(t)$ possesses no unique sign near
$t_1$. Then we have to prove proposition \ref{prop.first_estimates} and the two
subsequent lemmas \ref{lem.dx} and \ref{lem.dt}. The discussion at the
beginning of this section has shown, that some of these quantities are
already bounded. The crucial step is, to find a bound for lapse and its
derivatives.

The idea is to reparametrize the foliation as has been done for a CMC
foliation in \cite{br}. The effect of the reparametrization on lapse
and shift was outlined in section \ref{s.spacetimes}. Let us introduce a
function $\tau$ by 
\begin{equation*}
   \tau := t + \int_{t_1}^{t} N(u,x_0)\,du
   \komma
\end{equation*}
where $\int_{t_1}^{t}$ means $\lim_{s \rightarrow t_1} \int_s^t$. It is well
defined, since $\int N(\gamma(u))\,du$ along the 
integral curves of the normals of the leaves measures the length of
$\gamma$, which we already know to be finite, and $\abs{N'} \le C$ ensures
the integrability of $N(t,x_0) \le N(\gamma(t)) + C$. This construction
works in either case, whether $N$ is bounded or diverges uniformly to
infinity. The function 
$t \longmapsto \tau(t)$ is monotone since 
$\tfrac{d\tau}{dt} = 1 + N(t,x_0)$ and turns out to be an orientation
preserving diffeomorphism 
$]t_1,t_2[\; \longrightarrow \;]\tau_1=t_1,\tau_2[\,$, thus
$\tau$ can be used as a new time function for the foliation, which squeezes
the time differences of neighbouring leaves compared with the normal
vector by adding to $t$ the normal component of the length of the
piece of the $x_0=\const$ line connecting $(t_1,x_0)$ with $(t,x_0)$.
This produces a new lapse function and a new shift vector:
\begin{gather*}
   \tilde{N} = \left( \frac{d\tau}{dt} \right)^{-1} N
             = \frac{N}{1+N(t,x_0)} \\
   \tilde{\nu} = \left( \frac{d\tau}{dt} \right)^{-1} \nu
               = \frac{\nu}{1+N(t,x_0)}
   \punkt
\end{gather*}
Conversely, $t$ stretches time differences of adjacent leaves, such that we
get the inverse transformation by 
$t = \tau - \int_{\tau_1}^{\tau} \tilde{N}(u,x_0)\,du$, which subtracts from
$\tau$ the former added 'length' of the $x_0=\const$ line now
expressed in terms of the new time coordinate. By the inverse
transformation, we see that
$\frac{dt}{d\tau}=1-\tilde{N}(\tau,x_0)$, and therefore get the relation
$1-\tilde{N}(\tau,x_0) = (1+N(t,x_0))^{-1}$ in points with spatial
coordinate $x_0$.

The benefit of this reparametrization is, that due to $\abs{N'} \le C$,
$\tilde{N}$ and $\abs{\tilde{N}'}$ remain bounded, and we could try to
analyse the field equations according to the reparametrized
foliation. Inspection of these equations shows, that we get the field
equations in the new coordinates from equations \eqref{e.ch}-\eqref{e.p'}
by replacing everywhere $\d_t$ by $\d_{\tau}$, $N$ by $\tilde{N}$, $\nu$ by
$\tilde{\nu}$ (since reformulating of the constraint and evolution
equations preserves their form) while the lapse equation
\eqref{e.l} and the PMC condition \eqref{e.p} depend on the
parametrization, and hence have to be modified: The subtraction of $A^2$ on
the right-hand side stems from the term $A^2 Nn(H)$, where the PMC condition
sets $Nn(H)$ equal to one. Replacing $N$ by $\tilde{N}$ yields 
$(Nn(H))(1+N(t,x_0))^{-1}=(1+N(t,x_0))^{-1}=1-\tilde{N}(\tau,x_0)$.
The PMC condition \eqref{e.p}, which holds in the old
parametrization, can be formulated in the new coordinates by expressing
$\nu$ in terms of $\tilde{\nu}$, which produces an additional summand
$1-\tilde{N}(\tau,x_0)$ on the right-hand side, besides the replacement of
$\nu$ by $\tilde{\nu}$.

Now we start to analyse the new field equations, trying to retain the same
line of thought of section \ref{s.estimates}: The analysis already done
here bounds $A,A^{-1},\abs{A'},a,a^{-1},\abs{H},\abs{K}$ as
well as, of course, $\tilde{N}$ and $\abs{\tilde{N}'}$. The arguments
carried out in section \ref{s.estimates} bound now 
$\abs{\tilde{\nu}},\abs{\tilde{\nu}'}$ and $\abs{H'}$.
Finally we need a bound for $\tilde{N}^{-1}$. Since it is not clear, in the
case where $N$ diverges near $t_1$, whether $N$ grows monotonically or not,
we divide the time interval $]t_1,t_2[$ into two
subsets. First consider all points, where $N(t,x_0) \ge M$, with some
suitable chosen real number $M$, which will be specified later. Then we get
the estimate 
\begin{equation*}
   \absbig{1-\tilde{N}(\tau,x)}
      = \absbig{1-\frac{N(t,x)}{1+N(t,x_0)}}
      = \absbig{\frac{1+N(t,x_0)-N(t,x)}{1+N(t,x_0)}}
      \le \frac{1+C}{1+M}
   \komma
\end{equation*}
where $C$ is an upper bound for $\abs{N(t,x_0)-N(t,x)}$, whose existence
is guaranteed by the bounds for $\abs{N'}$ and the volume of the leaves of
the foliation. Choose now $M=2C$ and we get 
$\abs{1-\tilde{N}} \le \frac{1+C}{1+2C} < 1$, thus $\tilde{N}$ is bounded
away from zero for all points, where $N(t,x_0) \ge M$. Consider now the
points, where $N(t,x_0) \le M$. Then we have
\begin{equation*}
   \absbig{1-\tilde{N}(\tau,x_0)}
      = \absbig{1-\frac{N(t,x_0)}{1+N(t,x_0)}}
      = \absbig{\frac{1+N(t,x_0)-N(t,x_0)}{1+N(t,x_0)}}
      \ge \frac{1}{1+M}
      \punkt
\end{equation*}
Now we can apply the arguments in \cite{r5} to the modified lapse equation,
as has been done in section \ref{s.estimates} (whereupon the factor
$1-\tilde{N}(\tau,x_0)$ causes no trouble, since it is bounded from above
and below) and we get a bound for $\tilde{N}^{-1}$ for all
points, where $N(t,x_0) \le M$, and we are done.\zeile
Therefore we have estimated all quantities appearing in proposition
\ref{prop.first_estimates}.
lemma \ref{lem.dx} is also true for the new field equations without the
need to modify the proof, while the
proof of lemma \ref{lem.dt} has to be modified, since the argument
bounding $\abs{\d_t^{m+1}\d_x^n \tilde{N}}$ does not carry over. 
The lapse equation and the equation for its time derivatives are
\begin{align*}
   (A\tilde{N}')' 
      & = A^3 \tilde{N} 
      \Big( \tfrac{1}{2}(H-K)^2 + K^2
             + 4\pi(\rho+\tr S) 
      \Big) 
      - A^3 \left( 1-\tilde{N}(\tau,x_0) \right) \\
   (A\d_t^{m}\tilde{N}')' 
      & = A^3 \d_t^{m}\tilde{N} 
      \Big( \tfrac{1}{2}(H-K)^2 + K^2
            + 4\pi(\rho+\tr S) 
      \Big) \\
      & - A^3 \left( 1-\d_t^{m}\tilde{N}(\tau,x_0) \right) + B
   \komma
\end{align*}
where $B$ denotes an already bounded quantity at the corresponding stage in
the inductive argument in lemma \ref{lem.dt}. In this situation 
lemma 1 in \cite{br} applies literally, bounding the time derivatives of
the lapse function, therefore completing the proof of lemma
\ref{lem.dt}.\zeile 
Putting all arguments together we are able to extend the foliation beyond
$t_1$, contradicting the assumed maximality of the interval of existence,
thus arriving at the following
\begin{theorem} \label{thm.pmcsphere} \ein
   Let $(M,g,f)$ be a surface symmetric solution of the
   Einstein-Vlasov system with spherical symmetry.\zeile
   Then we can foliate the whole spacetime by a PMC 
   foliation, where the time function takes on all real values and the mean
   curvature of the leaves tends uniformly to $\pm \infty$ for 
   $t \rightarrow \pm\infty$, thus producing crushing singularities.
\end{theorem}
This theorem provides barrier surfaces (see \cite{g}), establishing now the
existence of CMC hypersurfaces for each value of the mean curvature,
therefore proving the closed universe recollapse conjecture in this case:
\begin{corollary}\label{cor.cmcsphere}\ein
   In the situation of theorem \ref{thm.pmcsphere}, the spacetime possesses
   a global CMC foliation and the mean curvature takes on all real
   values. In particular, the spacetime admits a maximal Cauchy surface.
\end{corollary}
\subsubsection{Spacetimes with plane and hyperbolic symmetry}
\label{ss.planehyp}
We will see, that most of the arguments performed in the spherically
symmetric case carry over
to the past domain of dependence $D^-(\Sigma)$ in the expanding models
(compare \ref{ss.expanding} for a precise definition of this terminology). Due
to this fact we perform the following analysis only on the half-open
time interval $]t_1,0]$ and assume
further, $(M,g)$ to be non-flat in the plane symmetric case and the mass
function \eqref{e.mass} to be positive on $\Sigma$ in the hyperbolic
case.\zeile
Again we can exclude some cases, already investigated (although this does
not matter, since the arguments here apply to these cases). 
It has been proven in \cite{r4} and \cite{br}, that given 
a symmetric Cauchy surface with negative constant mean curvature $H_0$
(remember the restrictions on the mean curvature in the expanding models,
compare \ref{ss.expanding}), there exists a global CMC foliation with the mean
curvature taking all values in the interval $]-\!\infty,H_0]$.
So we can assume without loss of generality, that the mean curvature $H$ on
$\Sigma$ is not constant and not everywhere positive. Again, 
this assumption ensures the existence of a local in time PMC foliation
$]t_1,t_2[\,\times \Sigma$ of a neighbourhood of $\Sigma$ in $M$.

Now we find similar a priori bounds
\begin{itemize}
\item In $D^-(\Sigma)$ $r$ is bounded, since $dr$ is future pointing
   timelike everywhere and $\Sigma$ is compact. $\mass^{-1}$ is bounded on
   $\Sigma$ by assumption in 
   the hyperbolic case and in the plane symmetric case this follows from
   lemma 2.4 in \cite{r4}. By the mass flux equation \eqref{e.massflux}
   together with the non-negative pressures condition, $\mass$ grows
   along past 
   pointing timelike curves, thus $\mass^{-1}$ is bounded in all of
   $D^-(\Sigma)$. 
   Theorem 2.1 in \cite{b} then shows, that all
   timelike curves in $D^-(\Sigma)$ have finite length.
\item Applying the arguments in the corresponding place of section
   \ref{ss.spherically} we get bounds for 
   $r^{-1}$ and $\mass$ for any finite time interval of the form $]t_1,0]$.
\item Again, the corresponding argument performed in the spherically
   symmetric case holds, and we get upper and lower bounds for the volumes
   of arbitrary Cauchy surfaces in $M$.
\end{itemize}

Finally, the volume of the leaves $S_t$ is given by 
$V(t)=C a^{-1} \int_{S^1} r^3$, establishing bounds for the first
fundamental form and its inverse (in $D^-(\Sigma)$) as has been shown in
\ref{ss.spherically}, therefore all of the remaining arguments performed
there apply and we get the  
\begin{theorem} \label{thm.pmcplane}  \ein
   Let $(M,g,f)$ be a surface symmetric solution of the
   Einstein-Vlasov system with plane or hyperbolic symmetry and $\Sigma$
   a symmetric Cauchy surface in $M$.\zeile
   If $(M,g)$ is non-flat in the plane symmetric case and the mass function is
   positive on $\Sigma$ in the hyperbolic case, then we can foliate all of
   the past of $\Sigma$ by PMC hypersurfaces, where the time function takes
   on all values in the interval $]-\!\infty,0]$ and the mean
   curvature of the leaves tends uniformly to $-\infty$ for 
   $t \rightarrow -\infty$.
\end{theorem}
Using again the PMC leaves as barrier surfaces we get the
\begin{corollary}\label{cor.cmcplane}\ein
   In the situation of theorem \ref{thm.pmcplane} $D^-(\Sigma)$ possesses a
   CMC Cauchy surface for each value of the mean curvature in
   $]-\!\infty,\min_{\Sigma}H[$. 
\end{corollary}
\subsubsection{Comparing the results}
To close this chapter, there are some remarks concerning the 
theorems \ref{thm.pmcsphere} and \ref{thm.pmcplane} and the relation to
the work done in the preceding sections.

At first, it is obvious, that theorem \ref{thm.pmcsphere} generalizes
theorem \ref{thm.ev} in the spherically symmetric case. In the plane and
hyperbolic case theorem \ref{thm.pmcplane} does not generalize the theorems
\ref{thm.ev}, \ref{thm.em} and \ref{thm.evm}. The latter ones only
establish a global PMC 
foliation unless the mean curvature of the leaves becomes zero somewhere
(and again the leaves provide barrier surfaces for CMC Cauchy surfaces,
establishing corollary \ref{cor.cmcplane} in the situation of those
theorems). Theorem \ref{thm.pmcplane} is not restricted to this condition,
but it needs the extra non-negative pressures condition, which excludes
electromagnetic fields. 

Apart from the non-negative pressures condition the assumption of positive
mass on $\Sigma$ is a non-trivial constraint in the hyperbolic case, while
automatically fulfilled in the non-flat plane symmetric case by lemma 2.4
in \cite{r4}. As shown above, the positivity of mass is needed to obtain a
bound for the length of timelike curves in $M$, by applying the arguments
in \cite{b}, thus necessary to the construction done here.
%
%
%
%
%
%

%%% Local Variables: 
%%% mode: latex
%%% TeX-master: "main"
%%% End: 

%
%
%
%
%
%
\section{Conclusion and outlook}
First I list the main results achieved in this work
\begin{itemize}
\item For the spacetimes considered so far the existence
   of a global PMC foliation has been shown, where
   several matter models have been taken into account.
\item For the model, which is not expanding or contracting everywhere, the
   closed universe recollapse conjecture has been proved. In particular the
   foliation covers the whole Cauchy development of the initial
   hypersurface, with a crushing singularity both in the distant past and
   future. Moreover the spacetime possesses a maximal hypersurface.
\item In the expanding models the foliation covers at least the whole past
   of the initial hypersurface towards a crushing singularity.
\end{itemize}
The choice of matter turned out to be important only in so far, as some
energy conditions are satisfied and the matter fields do not become
singular in a regular geometric background.\zeile
The necessary energy conditions are the dominant and strong energy
conditions. Furthermore to
obtain stronger results, which do not rely on a fixed sign condition for
the mean curvature, the non-negative pressures condition is required, to
ensure that the Hawking mass does not tend to zero, contributing enough
attraction, that the lifetime of timelike curves become finite.\zeile
The rather strong results about locally spatially homogeneous spacetimes in
\cite{r3} likewise rely on this condition, which appears in the slightly
relaxed form, that only the sum of the pressures has to be non-negative,
due to the high degree of symmetry in those models. This relaxed condition
even permits electromagnetic fields, a type of matter, which does not
satisfy the non-negative pressures condition, thus not leading to the
stronger results in the context of spacetimes with less symmetry. 

The close analogy of the proofs for the global existence of CMC and PMC
foliations indicates, that all results obtained for CMC foliations may also
be proved for PMC foliations. If this conjecture turns out to be true (the
second part of this paper will give another positive example for this
conjecture), then
we are no longer concerned with the topological restrictions imposed on the
existence of CMC hypersurfaces, having a much more flexible tool at
hand. Moreover, the global results on PMC foliations can be used to provide
barrier surfaces, which guarantee the existence of a CMC foliation, where
the mean curvature ranges between these barriers (compare the corollaries
\ref{cor.cmcsphere} and \ref{cor.cmcplane}).

The results demonstrate, that in the cases considered in this work a
satisfying answer has been given to the global existence problem for PMC
foliations. Then the related question arises, whether the foliation covers
the whole spacetime. The present work gives only a partial answer to this
question. Denoting the initial Cauchy surface by $\Sigma$, we saw, that we
have covered the whole past $D^-(\Sigma)$ in the expanding spacetimes and
the whole Cauchy development $D(\Sigma)$ in the recollapsing models. These
positive results have two limits
\begin{enumerate}
\item In the expanding models there remains an open question about the
   future of $\Sigma$ in $M$. Either the mean curvature becomes zero
   somewhere or there is no control on the radius function towards
   the expanding direction. In each case the present techniques do not 
   apply. In addition there are topological obstructions in the expanding
   direction, preventing the mean curvature to become positive, a foliation
   of the future of $\Sigma$ must stop before this happens. But this does
   not imply, that the whole of $D^+(\Sigma)$ can be covered by such a
   foliation, since the leaves of the foliation may become null or
   non-compact, where the general notion of singularity avoidance comes
   up.\zeile
   Namely, compared with the strong results obtained in \cite{r3} in the
   locally spatially homogeneous models, we require more information about
   the asymptotic behaviour of the foliation to obtain results on geodesic
   completeness. 
\item In the contracting direction and in the recollapsing models things
   look different. In the contracting direction the
   theorems ensure the existence of crushing singularities, and all of
   the Cauchy development of $\Sigma$ is covered by the foliation. This is
   a consequence of Hawking's famous incompleteness theorem for globally
   hyperbolic spacetimes, 
   satisfying the timelike convergence condition, where the maximal
   time of existence is estimated by $3/\abs{H}$ and 
   $H \rightarrow -\infty$. But there is no information about the boundary
   of $D(\Sigma)$, which might be either a curvature singularity or merely
   a Cauchy horizon.\zeile
   The work in \cite{r3} indicates roughly, what remains to do. Relating
   the particle current density to the energy density and investigating the
   asymptotic behaviour of the former may give rise to unbounded
   curvature and produce a curvature singularity.
\end{enumerate}

For the spacetimes considered here, the work \cite{re} of Rein is
important. He 
succeeded in finding satisfying answers to the above questions, by
using a time function intimately tied to the symmetry. Unfortunately it is
not clear how to generalize his approach to other spacetimes.
Although the estimates done here also exploit the high symmetry of the
models, the construction itself does not depend on it, motivating this
work. A generalization
to spacetimes with two commuting local Killing vector fields will be the
content of the second part.
%
%
%
%
%

%%% Local Variables: 
%%% mode: latex
%%% TeX-master: "main"
%%% End: 

%
%
%
%
%
%
\pagebreak

%
%
%
%
%
%%% Local Variables: 
%%% mode: latex
%%% TeX-master: "main"
%%% End: 

%
%
%
%
%
\end{document}